\documentclass[pre,twocolumn,amsmath,amssymb,floatfix]{revtex4}

\usepackage{graphicx}
\usepackage{dcolumn} 
\usepackage{bm}      
\usepackage{color}

\begin{document}

\title{The decay of turbulence in rotating flows}
\author{Tomas Teitelbaum$^{1}$ and Pablo D.~Mininni$^{1,2}$}
\affiliation{$^1$ Departamento de F\'\i sica, Facultad de Ciencias 
Exactas y Naturales, Universidad de Buenos Aires, IFIBA and CONICET, 
Ciudad Universitaria, 1428 Buenos Aires, Argentina. \\
$^2$ NCAR, P.O. Box 3000, Boulder, Colorado 80307-3000, U.S.A.}
\date{\today}

\begin{abstract}
We present a parametric space study of the decay of turbulence in 
rotating flows combining direct numerical simulations, large eddy 
simulations, and phenomenological theory. Several cases are considered: 
(1) the effect of varying the characteristic scale of the initial 
conditions when compared with the size of the box, to mimic ``bounded'' 
and ``unbounded'' flows; (2) the effect of helicity (correlation between 
the velocity and vorticity); (3) the effect of Rossby and Reynolds 
numbers; and (4) the effect of anisotropy in the initial conditions. 
Initial conditions include the Taylor-Green vortex, the 
Arn'old-Beltrami-Childress flow, and random flows with large-scale 
energy spectrum proportional to $k^4$. The decay laws obtained in the 
simulations for the energy, helicity, and enstrophy in each case can be 
explained with phenomenological arguments that consider separate decays 
for two-dimensional and three-dimensional modes, and that take into 
account the role of helicity and rotation in slowing down the energy 
decay. The time evolution of the energy spectrum and development of 
anisotropies in the simulations are also discussed. Finally, the effect 
of rotation and helicity in the skewness and kurtosis of the flow is 
considered.
\end{abstract}
\maketitle

\section{Introduction}

Nature presents several examples of rotating flows. Rotation influences 
large-scale motions in the Earth's atmosphere and oceans, as well as 
convective regions of the sun and stars. Rotation is also important in 
many industrial flows, such as turbo machinery, rotor-craft, and 
rotating channels. In a rotating system, the Coriolis force, linear in 
the velocity, modifies the flow nonlinear dynamics when strong enough. In 
its presence, the Navier-Stokes equation becomes a multi-scale problem with 
a ``slow'' time scale $\tau_L \sim L/U$ associated with the eddies at a 
characteristic scale $L$ ($U$ is a characteristic velocity), and a ``fast'' 
time scale $\tau_\Omega \sim 1/\Omega \sim \tau_L Ro$ associated with 
inertial waves. The dimensionless Rossby number $Ro$ is the ratio of 
advection to Coriolis forces, and measures the influence of rotation upon 
the nonlinear dynamics of the system (decreasing as rotation becomes 
dominant).

Resonant wave theory provides a framework to study the effect of rapid 
rotation in turbulence 
\cite{Greenspan1968,Greenspan1969,Cambon1989,Waleffe1993,Cambon1997}. 
The separation between the fast and slow time scales results in a 
selection of the resonant triadic interactions as the ones responsible 
for the energy transfer among scales. As a result, energy transfer 
and dissipation is substantially decreased in the presence of 
strong rotation \cite{Cambon1989}. The resonant condition is also 
responsible for the transfer of energy towards two-dimensional (2D) slow 
modes, driving the flow to a quasi-2D state \cite{Cambon1989,Waleffe1993} 
(this result is often referred to in the literature as the 
``dynamic Taylor-Proudman theorem'', see e.g., \cite{Chen2005}). The 
development of anisotropy and reduction of the energy transfer and 
dissipation rates has been verified in numerical simulations 
\cite{Jacquin90,Bartello1994,Muller2007,Bokhoven2008,Mininni09} and 
experiments \cite{Morize2005,Morize2006}.

Similar arguments (see, e.g., \cite{Babin1996}) indicate that in the 
limit of fast rotation (small Rossby number) the slow 2D modes 
decouple from the remaining fast three-dimensional (3D) modes, and 
evolve under their own autonomous dynamics. Moreover, in that limit the 
averaged equation for the slow modes splits into a 2D Navier-Stokes 
equation for the vertically-averaged horizontal velocity and a passive 
scalar equation for the vertically-averaged vertical velocity. 
Although simulations of forced rotating flows \cite{Chen2005} and of 
ideal truncated rotating flows \cite{Bourouiba2008,Mininni2011} using 
periodic boundary conditions show good agreement with these predictions 
for small enough Rossby numbers, for long times the decoupling of slow 
and fast modes seems to break down. Also, some authors \cite{Cambon2004} 
argue that in unbounded domains no decoupling is achievable even for 
$Ro=0$.

Of particular importance for many geophysical and astrophysical problems 
is how turbulence decays in time. The problem is also important for 
laboratory experiments, as it provides, e.g., one way to measure changes 
in the energy dissipation rate associated with the presence of rotation. 
Even in the absence of rotation, the decay of isotropic turbulence 
proves difficult to tackle because of the different decay laws 
obtained depending on boundary and initial conditions. As an example, 
for bounded flows (i.e., flows for which the initial characteristic 
length is close to the size of the vessel) the energy decays as 
$\sim t^{-2}$ (see, e.g., \cite{Borue1995,Morize2006,Teitelbaum2009}). For 
unbounded flows (i.e., flows in an infinite domain, or in practice, flows 
for which the initial characteristic length is much smaller than the size 
of the vessel) a $\sim t^{-10/7}$ \cite{kolmo1941,Ishida2006} or 
$\sim t^{-6/5}$ \cite{Saffman1967,Saffman1967b,Morize2006} decay law is 
observed depending on whether the initial energy spectrum at large scales 
behaves as $\sim k^4$ or $\sim k^2$, respectively.

In the presence of rotation, the decay of turbulence becomes 
substantially richer, with decay rates depending not only on whether 
turbulence is bounded or unbounded and on the initial spectrum at 
large scales, but also, e.g., on the strength of background rotation 
(see \cite{Yang2004}). A detailed experimental study of this 
dependence can be found in Refs. \cite{Morize2005,Morize2006}, 
where the authors studied the energy decay of grid-generated turbulence 
in a rotating tank using particle image velocimetry, and found different 
decay laws depending upon the rate of rotation and the saturation (or 
not) of the characteristic size of the largest eddies. For large Rossby 
number they reported a decay $\sim t^\alpha$ with exponent 
$\alpha \approx -1.1$ for non rotating turbulence (consistent with the 
value of $-6/5$ predicted in \cite{Saffman1967,Saffman1967b} for $\sim k^2$ 
unbounded turbulence), which later turned to $\alpha \approx -2$ after the 
largest eddies grew to the experiment size. For small Rossby number this 
energy decay rate became smaller saturating at $\approx -1$. Similar 
results were reported in simulations in \cite{Bokhoven2008}, were a 
decrease from $\approx -10/7$ to $-5/7$ was observed as rotation was 
increased. These results are consistent with the reduction of the energy 
transfer discussed above. They also reported a steeper energy spectrum 
together with positive vorticity skewness for the rotating flows, and 
anisotropic growth of integral scales (see also 
\cite{Jacquin90,Squires1994}).

The decay rate of rotating turbulence also seems to depend on the 
helicity content of the flow. Helicity (the alignment between velocity 
and vorticity) is an ideal invariant of the equations of motion (a 
quantity conserved in the inviscid limit) with intriguing properties. 
In the absence of rotation, the presence of helicity does not modify 
the energy decay rate \cite{Polifke1989,Morinishi2001,Teitelbaum2009} nor 
the dissipation rate \cite{Rogers1987}. However, in the presence of 
rotation \cite{Morinishi2001} reported a further decrease of the energy 
transfer when both rotation and helicity were present, and 
\cite{Teitelbaum2009} showed that the decay rate of bounded rotating 
flows changes drastically depending on whether helicity is present or 
not.

In this paper we conduct a detailed study of parameter space of 
rotating helical flows, taking into account (1) the effect of varying 
the characteristic scale of the initial conditions when compared with 
the size of the box, (2) the effect of helicity, (3) the effect of 
Rossby and Reynolds numbers, and (4) the effect of anisotropy in the 
initial conditions. The numerical study uses a two pronged approach 
combining direct numerical simulations (DNS) and large eddy simulations 
(LES). Several initial conditions are considered, although when the 
characteristic initial scale of the flow is smaller than the size of 
the domain, we focus only on the case with large scale initial energy 
spectrum $\sim k^4$. The different decay laws obtained (which in some 
cases coincide with previous experimental or numerical observations, 
while in others are new) are explained using phenomenological arguments, 
and we classify the results depending on the relevant effects on each 
case.

After studying the decay laws, we study the evolution of anisotropy, 
of skewness and kurtosis, and the formation of columnar structures in 
the flow. We consider how helicity affects the evolution of skewness 
and kurtosis, and associate peaks observed in the time evolution of 
these quantities with the dynamics of the columnar structures.

The structure of the paper is as follows. In Sec. \ref{sec:equations} we 
introduce the equations, describe the DNS and LES, and give details of 
the initial conditions and different parameters used. Section 
\ref{sec:phenom} presents phenomenological arguments to obtain decay 
laws in turbulent flows with and without rotation. The phenomenological 
predictions are then compared with the numerical results in Sec. 
\ref{sec:decay}, which presents the energy, helicity, and enstrophy decay 
in all runs. The spectral evolution and development of anisotropy in the 
flows is discussed at the end of that section. The effect of initial 
anisotropy in the decay is considered in Sec.~\ref{sec:anisotropy}. A 
statistical analysis including evolution of skewness and kurtosis is 
presented in Sec.~\ref{sec:statistics}. Finally, Sec.~\ref{sec:conclusions} 
gives our conclusions.

\section{\label{sec:equations}Equations and models}

\subsection{Equations}

The evolution of an incompressible fluid in a rotating frame is described 
by the Navier-Stokes equation with the Coriolis force,
\begin{equation}
\partial_t {\bf u} + \mbox{\boldmath $\omega$} \times
    {\bf u} + 2 \mbox{\boldmath $\Omega$} \times {\bf u}  =
    - \nabla {\cal P} + \nu \nabla^2 {\bf u} ,
\label{eq:momentum}
\end{equation}
and the incompressibility condition,
\begin{equation}
\nabla \cdot {\bf u} = 0 ,
\label{eq:incompressible}
\end{equation}
where ${\bf u}$ is the velocity field, 
$\mbox{\boldmath $\omega$} = \nabla \times {\bf u}$ is the vorticity, the 
centrifugal term is included in the total pressure per unit of mass 
${\cal P}$, and $\nu$ is the kinematic viscosity (uniform density is 
assumed). The rotation axis is in the $z$ direction, so 
$\mbox{\boldmath $\Omega$} = \Omega \hat{z}$, where $\Omega$ is the 
rotation frequency.

As mentioned in the introduction, these equations are solved numerically 
using two different methods: DNS, and LES using a dynamical subgrid-scale 
spectral model of rotating turbulence that also takes into account the 
helicity cascade. All simulations were performed in a three dimensional 
periodic box of length $2 \pi$, using different spatial resolutions 
ranging from $96^3$ grid points for the lowest resolution LES runs up to 
$512^3$ for the highest resolution DNS.

\begin{table*}
\caption{\label{table:runs}Parameters used in the simulations: kinematic 
viscosity $\nu$, rotation rate $\Omega$, Reynolds number $Re$, Rossby 
number $Ro$, micro-Rossby number $Ro^{\omega}$, initial relative helicity 
$h$, relative helicity at the time of maximum dissipation $h^*$, and time 
of maximum dissipation $t^*$. The values of $Re$, $Ro$, and $Ro^{\omega}$ 
are always given at $t^*$. The last column succinctly describes the initial 
energy spectrum $E(k)$: the power law followed by the spectrum, the range 
of scales where this power law is satisfied, and the flow (TG for 
Taylor-Green, ABC for Arn'old-Beltrami-Childress, and RND for random).}
\begin{ruledtabular}
\begin{tabular}{lccccccccc}
Run&$\nu$&$\Omega$&$Re$&$Ro$&$Ro^\omega$&$h$&$h^*$&$t^*$& Initial $E(k)$ \\
\hline
D256-1 &$1.5\times 10^{-3}$& $0$ & $450$  & $\infty$
  & $\infty$ & $0$ & $9 \times 10^{-10}$ & $1.26$ 
  & $k^{-4}$ (4-14) TG \\
D256-2 &$1.5\times 10^{-3}$& $4$ & $550$  & $0.12$ & $1.28$
  & $0$ & $-1 \times 10^{-8}$ & $1.06$ 
  & $k^{-4}$ (4-14) TG \\
D256H-1 &$1.5\times 10^{-3}$ & $0$ & $600$  & $\infty$ & $\infty$
  & $0.95$ & $0.34$ & $2.28$ &$k^{-4}$ (4-14) ABC       \\
D256H-2 &$1.5\times 10^{-3}$& $4$ & $830$  & $0.08$ & $0.80$ 
  & $0.95$ & $0.65$ & $2.25$ & $k^{-4}$ (4-14) ABC      \\
D512-1 &$7\times 10^{-4}$& $4$ & $1100$ & $0.12$ & $1.82$ 
  & $0$ & $7 \times 10^{-9}$ & $0.88$ 
  & $k^{-4}$ (4-14) TG \\
D512-2 &$8.5 \times 10^{-4}$& $0$ & $420$  & $\infty$ & $\infty$
  & $8 \times 10^{-5}$ & $8 \times 10^{-4}$ & $0.60$ & $k^4$ (1-14) RND  \\
D512-3 &$8.5\times 10^{-4}$& $10$ & $450$  & $0.10$ & $0.95$
  & $4 \times 10^{-3}$ & $4 \times 10^{-3}$ &$0.70$ & $k^4$ (1-14) RND   \\
D512H-1 &$7\times 10^{-4}$& $4$ & $1750$ & $0.08$ & $1.15$ 
  & $0.95$ & $0.44$ & $1.70$ & $k^{-4}$ (4-14) ABC      \\
D512H-2 &$8 \times 10^{-4}$& $0$ & $440$ & $\infty$ & $\infty$
  & $0.90$ & $0.38$ & $0.94$ & $k^4$ (1-14) RND         \\
D512H-3 &$8 \times 10^{-4}$& $10$ & $530$  & $0.07$ & $0.70$ 
  & $0.99$ &$0.5$ & $1.50$ & $k^4$ (1-14) ABC           \\
L96-1 &$8.5\times 10^{-4}$& $0$ & $550$ & $\infty$ & $\infty$
  & $0.03$ &$0.02$ & $0.30$ & $k^4$ (1-14) RND          \\
L96-2 &$8.5\times 10^{-4}$& $2$ & $540$ & $0.42$ & $2.90$
  & $-0.03$ &$-0.02$ & $0.30$ &$k^4$ (1-14) RND         \\
L96-3 &$8.5\times 10^{-4}$& $4$ & $540$ & $0.21$ & $1.45$ 
  & $-0.03$ &$-0.02$ & $0.30$ &$k^4$ (1-14) RND         \\
L96-4 &$8.5\times 10^{-4}$& $6$ & $550$ & $0.14$ & $0.95$ 
  & $-0.03$ &$-0.02$ & $0.30$ &$k^4$ (1-14) RND         \\
L96-5 &$8.5\times 10^{-4}$& $8$ & $550$ & $0.11$ & $0.73$ 
  & $-0.03$ &$-0.02$ & $0.30$ &$k^4$ (1-14) RND         \\
L96-6 &$8.5\times 10^{-4}$& $10$ & $530$ & $0.08$ & $0.65$ 
  & $0.03$ &$0.02$ & $0.30$ &$k^4$ (1-14) RND           \\
L96H-1 &$8\times 10^{-4}$& $0$ & $500$ & $\infty$ & $\infty$
  & $0.90$ &$0.51$ & $0.70$ & $k^4$ (1-14) RND          \\
L96H-2 &$8.5\times 10^{-4}$& $10$ & $540$ & $0.08$ & $0.63$
  & $0.90$ &$0.70$ & $0.70$ & $k^4$ (1-14) RND          \\
L96H-3 &$8.5\times 10^{-4}$ & $10$ & $490$ & $0.08$ & $0.60$  
  & $0.99$ & $0.80$ & $1.15$ & $k^4$ (1-14) ABC         \\
L192-1 &$2\times 10^{-4}$& $0$ & $1200$ & $\infty$ & $\infty$
  & $-7\times 10^{-3}$ &$-6\times 10^{-3}$ & $0.10$ & $k^4$ (1-30) RND   \\
L192-2 &$2\times 10^{-4}$& $10$ & $1100$ & $0.22$ & $1.65$
  & $-7\times 10^{-3}$ & $-6\times 10^{-3}$ & $0.13$ & $k^4$ (1-30) RND  \\
L192H-1 &$2\times 10^{-4}$& $0$ & $950$ & $\infty$ & $\infty$
  & $0.90$ & $0.60$ & $0.30$ & $k^4$ (1-30) RND         \\
L192H-2 &$2\times 10^{-4}$& $10$ & $1000$ & $0.20$ & $1.60$
  & $0.94$ & $0.71$ & $0.38$ & $k^4$ (1-30) ABC         \\
L192HA-1 &$2\times 10^{-4}$& $10$ & $1200$ & $0.16$ & $1.40$
  & $0.90$ & $0.56$ & $0.50$ & $k^4$ (1-25) RND         \\
L192HA-2 &$2\times 10^{-4}$& $10$ & $1300$ & $0.14$ & $1.35$
  & $0.90$ & $0.59$ & $0.46$ & $k^4$ (1-25) RND         \\
L192HA-3 &$2\times 10^{-4}$& $10$ & $1300$ & $0.15$ & $1.35$
  & $0.90$ & $0.58$ & $0.45$ & $k^4$ (1-25) RND
\label{tab:paratable}
\end{tabular}
\end{ruledtabular}
\end{table*}

\subsection{Models}

In a DNS, all spatial and time scales (up to the dissipation scale) are 
explicitly resolved. The simulations were performed using a parallelized 
pseudo-spectral code \cite{Gomez1,Gomez2} with the two-third rule for 
dealiasing. As a result, the maximum wave number resolved in the DNS is 
$k_{max}=N/3$ where $N$ is the linear resolution; to properly resolve the 
dissipative scales the condition $k_{\eta}/k_{max} < 1$ must be satisfied 
during all simulations, where $k_{\eta}$ is the dissipation wave number. 
In practice, this condition is more stringent if reliable data 
about velocity gradients and high-order statistics of the flow are needed 
(see, e.g., \cite{Jimenez93}, where they indicate $k_{\eta}/k_{max}< 0.5$ 
for such studies).

The dissipation wave number as a function of time was computed for 
all simulations in two different ways: as the Kolmogorov dissipation wave 
number for isotropic and homogeneous turbulence 
$k_{\eta} = (\epsilon/\nu^3)^{1/4} = (\langle\omega^2\rangle/\nu^2)^{1/4}$ 
(where $\epsilon$ is the energy dissipation rate and 
$\langle\omega^2\rangle$ the mean square vorticity), and as the wave 
number where the enstrophy spectrum peaks. The Kolmogorov dissipation 
wave number was found to be always larger than the wave number where 
dissipation peaks, and in the following we therefore only consider the 
Kolmogorov scale as it gives a more stringent condition on the resolution.

In all DNS discussed below, the ratio $k_{\eta}/k_{max}$ was 
$\le 0.7$ at the time of maximum dissipation ($t \approx 1$ to 
$t\approx 3$ depending on the simulation), $0.2$ to $0.5$ at $t\approx 10$ 
(when the self-similar decay starts in most runs), and monotonously 
decreases to values between $0.05$ to $0.2$ at $t\approx 100$. The 
spatial resolutions used were $256^3$ and $512^3$ grid points, and an 
explicit second-order Runge-Kutta method was used to evolve in time, with 
a Courant-Friedrichs-Levy (CFL) number smaller than one.

In the LES approach, only the large scales are explicitly resolved, while 
the statistical impact on the resolved scales of scales smaller than a 
cut-off scale is modeled with simplified equations. To this end we use 
the spectral model derived in \cite{Baerenzung2008} for isotropic helical 
and non-helical turbulence, and its extension to the rotating case 
\cite{Baerenzung2010}. The model is based on the eddy damped quasi-normal 
Markovian (EDQNM) closure to compute eddy viscosity and eddy noise, and 
assumes unresolved scales (scales smaller than the cut-off) are isotropic. 
Both eddy viscosity and noise are computed considering the contribution 
from the energy and the helicity spectra (see, e.g., \cite{Li06} for 
another subgrid model that takes into account the effect of helicity). The 
model adapts dynamically to the inertial indices of the resolved energy 
and helicity spectra, and as a result it is well suited to study rotating 
turbulence for which the scaling laws are not well known and may depend 
on the Rossby number. For a validation of the LES against DNS 
results, the reader is referred to \cite{Baerenzung2008,Baerenzung2010}.

The subgrid model starts by applying a spectral filter to the equations; 
this operation consists in truncating all velocity components at wave 
vectors ${\bf k}$ such that $|{\bf k}| = k > k_c$, where $k_c$ is the 
cut-off wave number. One then models the transfer between the large 
(resolved) scales and the small (subgrid unresolved) scales of the flow 
by adding eddy viscosity and eddy noise to the equations for the resolved 
scales. These are obtained solving the EDQNM equations for estimated 
energy and helicity spectra in the subgrid range. To this end, an 
intermediate range is defined, lying between $k_c'$ and $k_c$ (in most 
cases $k_c'=k_c/3$), where the energy spectrum is assumed to present a 
power-law behavior possibly followed by an exponential decrease. As an 
example, for the energy the following expression is used:
\begin{equation}
E(k,t) = E_0 k^{-\alpha_E} e^{-\delta_E k}, \quad k_c' \le k< k_c \ . 
\label{fit_Ev}
\end{equation}
The coefficients $\alpha_E$, $\delta_E$ and $E_0$ are computed at each 
time step doing a mean square fit of the resolved energy spectrum. The 
spectrum is extrapolated to the unresolved scales using these coefficients, 
and the EDQNM equations are solved. Then one solves the Navier-Stokes 
equation (\ref{eq:momentum}) with an extra term on the r.h.s.~which in 
spectral space takes the form
\begin{equation}
- \nu\left(k|k_c,t\right) k^2 {\bf u}({\bf k},t) ,
\end{equation}
where ${\bf u}({\bf k},t)$ is the Fourier transform of the velocity field 
${\bf u}({\bf x},t)$, $-k^2$ is the Laplacian in Fourier space, and 
$\nu\left(k|k_c,t\right)$ is an eddy viscosity proportional to the ratio 
of the so-called absorption terms in EDQNM to the energy (and helicity) 
spectrum. Eddy noise is added in a similar manner (for more details, see 
\cite{Baerenzung2008,Baerenzung2010}).

LES runs using this model have a resolution of either $96^3$ or $192^3$ 
grid points. A pseudo-spectral method is also used, but without dealiasing, 
resulting in the maximum wave number $k_{max}=N/2$. As in the DNS, an 
explicit second-order Runge-Kutta method is used to evolve in time. 

Parameters for all sets of runs are listed in Table \ref{table:runs}. DNS 
runs are labeled with a D, followed by the linear resolution, a letter 
``H'' if the run has helicity, a letter ``A'' if the initial energy 
spectrum is anisotropic, and the run number. LES runs start with an L, 
followed by numbers and letters using the same convention as in the DNS.

\subsection{Initial conditions and definitions}


As mentioned in the introduction, we are interested in the decay laws 
obtained in the system depending on properties of the initial conditions 
and the amount of rotation. In particular, we will vary the initial amount 
of helicity, the initial energy containing scale 
(with respect to the largest available scale in the box), the shape of 
the energy spectrum, and the strength of turbulence and of rotation as 
controlled by the Reynolds and Rossby numbers respectively.

Helicity is an ideal invariant of the Navier-Stokes equation which 
measures the alignment between velocity and vorticity. If zero, the 
initial conditions are mirror-symmetric, so it also measures the departure 
from a mirror-symmetric state. We define the net helicity as 
\begin{equation}
H=\left<{\bf u}\cdot\mbox{\boldmath $\omega$}\right>,
\end{equation}
where the brackets denote spatial average. We also define the relative 
helicity as
\begin{equation}
h = \frac{H}{\left<|{\bf u}||\mbox{\boldmath $\omega$}|\right>} ,
\end{equation}
which is bounded between $-1$ and $1$ and can be interpreted as the mean 
cosine of the angle between the velocity and the vorticity.

To control the net amount of relative helicity in the initial conditions 
we consider three different flows: a superposition of Taylor-Green (TG) 
vortices
\cite{Taylor37},
\begin{eqnarray}
{\bf u}_{TG} &=& U \sin(k_0x)\cos(k_0y)\cos(k_0z)\hat{x} - \nonumber \\
{} && U \cos(k_0x)\sin(k_0y)\cos(k_0z)\hat{y} ,
\label{eq:TG}
\end{eqnarray}
a superposition of Arn'old-Beltrami-Childress (ABC) flows \cite{Childress},
\begin{eqnarray}
{\bf u}_{ABC}&=& \left[B\cos(k_0y)+A\sin(k_0z)\right]\hat{x}+\nonumber \\
{} && \left[C\cos(k_0z)+A\sin(k_0x)\right]\hat{y} + \nonumber \\
{} && \left[A\cos(k_0x)+B\sin(k_0y)\right]\hat{z} ,
\label{eq:ABC}
\end{eqnarray}
(with $A=0.9$, $B=1.1$, and $C=1$), and a superposition of Fourier modes 
with random phases (RND) in which we use the algorithm described in 
\cite{Pouquet78} to control the relative helicity. In each case, the flows 
are superposed in a range of wave numbers as described in Table 
\ref{tab:paratable}, and with global amplitudes for each wave number to 
give the desired slope in the initial energy spectrum.


The TG vortex is non-helical ($h=0$) and has no energy in the $k_z=0$ 
modes, whose amplification in the rotating cases (see below) can thus be 
attributed only to a bidimensionalization process. The TG vortex was 
originally motivated as an initial condition which leads to rapid 
development of small spatial scales, and also mimics the von K\'arm\'an 
flow between two counter-rotating disks used in several experiments 
\cite{Lyon2007}. The ABC flow is an eigenfunction of the curl operator and 
as a result has maximum helicity ($h=\pm 1$ depending on the sign of 
$k_0$, when only one value of $k_0$ is excited), whereas the RND flow 
allows us to tune the amount of initial relative helicity between $-1$ 
and 1 as well as the initial anisotropy. 

When generating the flows, two different initial energy spectra were 
considered. To study initial conditions with characteristic length 
close to the size of the computational domain, a spectrum 
$E(k) \sim k^{-4}$ for $k\in [4,14]$ (followed by exponential decay) was 
imposed. To study initial conditions with length smaller than the domain 
size, we imposed a spectrum $E(k) \sim k^{4}$ for $k\in [1,14]$, $[1,25]$, 
or $[1,30]$ (also followed by exponential decay). In the latter case, the 
characteristic length can grow in time as the spectrum peaks around $k=14$, 
$25$, or $30$. This allows us to mimic (at least for a finite time before 
the characteristic length reaches the domain size) the decay of unbounded 
flows. The characteristic length will be associated in the following with 
the flow integral scale, which is defined as 
\begin{equation}
L = 2\pi \frac{\sum_k k^{-1} E(k)}{\sum_k E(k)} ,
\label{eq:intlength}
\end{equation}
where $E(k)$ is the isotropic energy spectrum.

Simulations in Table \ref{tab:paratable} are also characterized by 
different Reynolds and Rossby numbers. The Reynolds 
and Rossby numbers in the Table are defined as
\begin{equation}
Re = \frac{UL}{\nu} 
\label{eq:Re}
\end{equation}
and 
\begin{equation}
Ro = \frac{U}{2 \Omega L} ,
\label{eq:Ro}
\end{equation}
respectively. Of importance is also the micro-scale Rossby number 
(see e.g., \cite{Jacquin90})
\begin{equation}
Ro^\omega = \frac{\omega}{2 \Omega} ,
\label{eq:Row}
\end{equation}
which can be interpreted as the ratio of the convective to the Coriolis 
acceleration at the Taylor scale. The Rossby number $Ro$ must be small 
enough for rotation to affect the turbulence, while the micro-Rossby 
number $Ro^\omega$ must be larger than one for scrambling effects of 
inertial waves not to completely damp the nonlinear term, which would 
lead to pure exponential viscous energy decay \cite{Cambon1997}. In all 
runs in Table \ref{tab:paratable}, $Ro$ and $Ro^\omega$ are one order of 
magnitude apart at the time of maximum enstrophy $t^*$, and this interval 
is roughly preserved throughout the simulations.


Here and in the following, the isotropic energy spectrum is defined by 
averaging in Fourier space over spherical shells
\begin{equation}
E(k,t) = \frac{1}{2} \sum_{k\le |{\bf k}|<k+1} {\bf u}^*({\bf k},t)
\cdot {\bf u}({\bf k},t) ,
\label{eq:isoenespec}
\end{equation}
where ${\bf u}({\bf k},t)$ is the Fourier transform of the velocity 
field, and the asterisk denotes complex conjugate. Other two spectra 
can also be used to characterize anisotropy.

On the one hand, the so-called ``reduced'' energy spectra $E(k_\perp)$ 
and $E(k_\parallel)$ are defined averaging in Fourier space over 
cylinders and planes respectively. More specifically, the reduced 
energy spectra as a function of wave numbers $k_\perp$ with 
${\bf k}_\perp=(k_x,k_y,0)$, and $k_{\parallel}$ with 
${\bf k}_{\parallel}=(0,0,k_z)$, are defined by computing the sum 
above over all modes in the cylindrical shells 
$k_\perp\le |{\bf k}_\perp|<k_\perp+1$ and over planes 
$k_\parallel \le |{\bf k}_{\parallel}|<k_\parallel+1$ respectively 
(isotropic and reduced spectra for the helicity are defined in the same 
way). From the reduced spectra, perpendicular and parallel integral 
scales can be defined; e.g., for the perpendicular direction,
\begin{equation}
L_{\perp} = 2\pi \frac{\sum_{k_{\perp}} k_{\perp}^{-1} E(k_{\perp})}
    {\sum_{k_{\perp}} E(k_{\perp})} .
\end{equation}

On the other hand, more information of the spectral anisotropy can be 
obtained studying the axisymmetric energy spectrum $e(k_\parallel,k_\perp)$ 
(see, e.g., \cite{Cambon1989,Cambon1997}). Assuming the flow is 
axisymmetric, the three-dimensional spectrum can be integrated around the 
axis of rotation to obtain a spectrum that depends only on $k_\parallel$ 
and $k_\perp$, which relates to the reduced energy spectra as follows:
\begin{equation}
E(k_{\parallel})=\sum_{k_{\perp}}{e(k_\parallel,k_\perp)} ,
\end{equation}
and
\begin{equation}
E(k_{\perp})=\sum_{k_{\parallel}}{e(k_\parallel,k_\perp)}.
\end{equation}

\begin{figure}
\includegraphics[width=8cm,height=6cm]{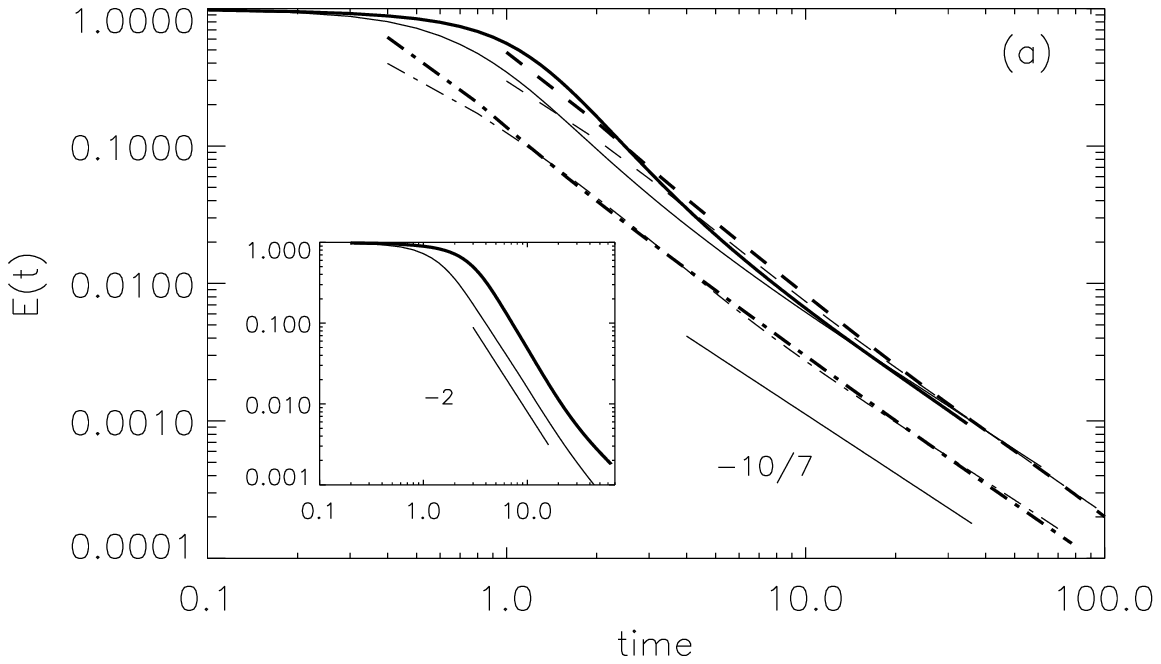}
\includegraphics[width=8cm,height=6cm]{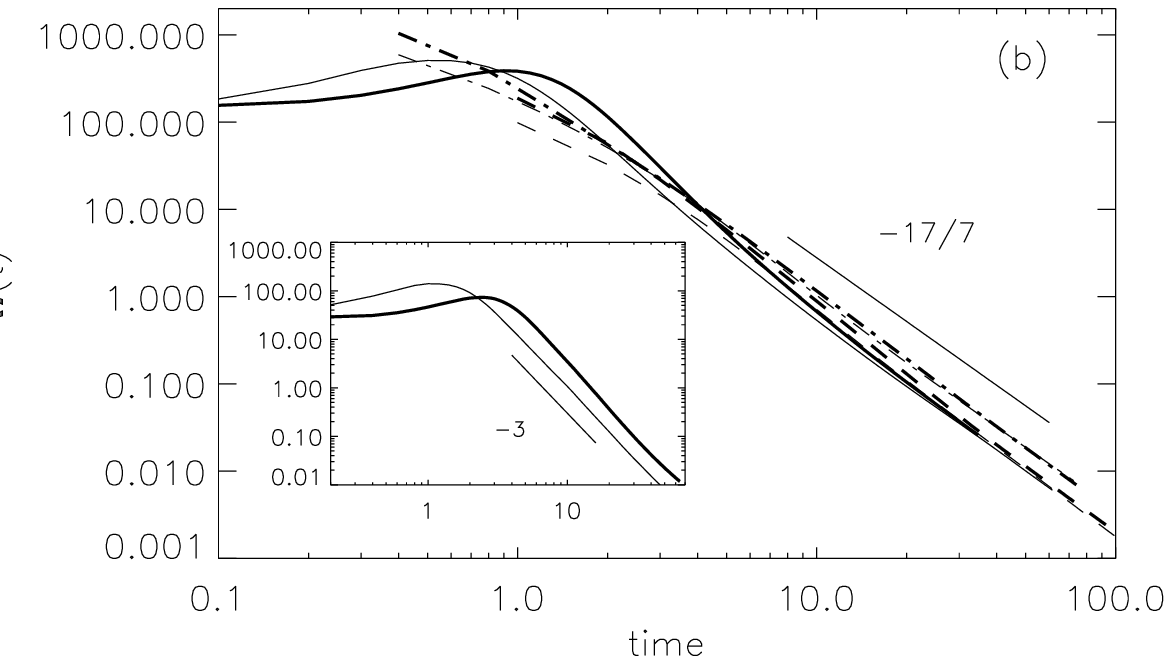}
\includegraphics[width=8cm,height=6cm]{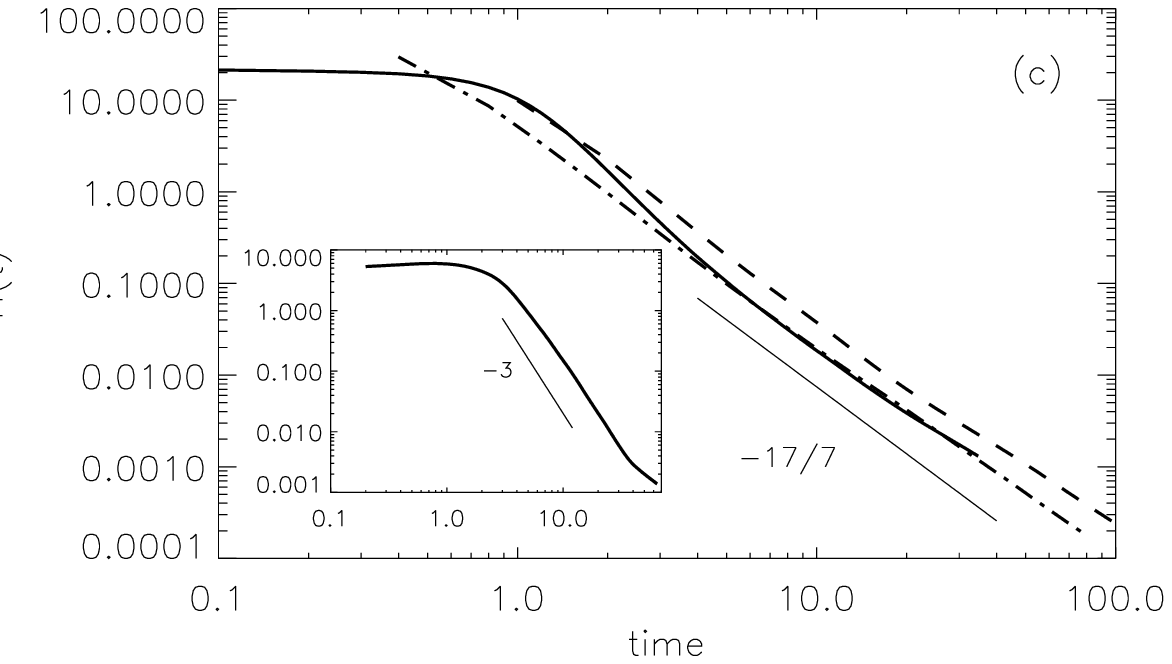}
\caption{(a) Energy decay for non-rotating unbounded runs. Non-helical runs 
D512-2 (solid), L96-1 (dashed), L192-1 (dash-dotted), and helical runs 
D512H-2 (solid, thick), L96H-1 (dashed, thick), and L192H-1 
(dash-dotted, thick) are shown. A $-10/7$ slope is shown as a reference. The 
inset shows the energy decay for non-rotating bounded runs D256-1 (solid), 
and D256H-1 (solid, thick). (b) Enstrophy decay for the same runs, with a 
$-17/7$ slope shown as a reference. The inset shows the enstrophy decay in 
the bounded runs. (c) Helicity decay in the unbounded helical runs of (a); 
the inset shows the helicity decay for the bounded helical run D256H-1.}
\label{fig:decays}
\end{figure}

\begin{figure}
\begin{center}
\includegraphics[width=9cm]{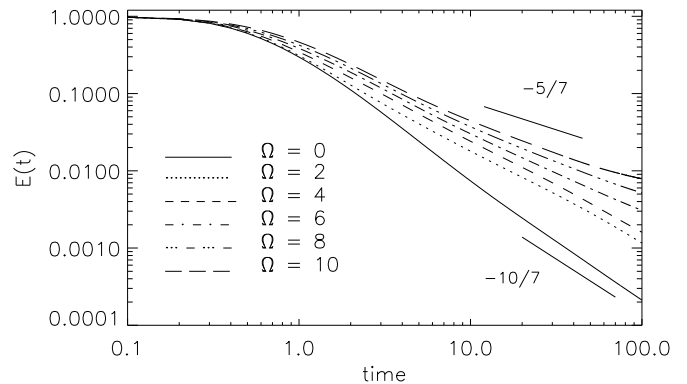}
\caption{Energy decay for different values of $\Omega$ from $0$ to $10$ 
for unbounded, non-helical runs L96-1, L96-2, L96-3, L96-4, L96-5, and 
L96-6. The decay becomes slower with increasing rotation rate. We also 
show $t^{-10/7}$ and $t^{-5/7}$ laws as references.}
\label{fig:decay_dif_rot_rates}
\end{center}
\end{figure}

\section{\label{sec:phenom}Time evolution - Phenomenology}

We present now phenomenological arguments that will become handy 
to understand the different decay rates that are observed in our simulations 
as well as in previous studies. Some of the arguments are well known, while 
others are new, and we quote previous derivations when needed.

\subsection{Non-rotating flows}

\subsubsection{Bounded}
From the energy balance equation
\begin{equation}
\frac{dE}{dt} \sim \epsilon
\label{eq:balance}
\end{equation}
where $\epsilon$ is the energy dissipation rate, Kolmogorov phenomenology 
leads to 
\begin{equation}
\frac{dE}{dt} \sim \frac{E^{3/2}}{L} ,
\label{eq:kolmodecay}
\end{equation}
where $E=E(t)\sim kE(k)$ and $L$ is an energy containing length scale. For 
bounded flows where $L\sim L_0$ ($L_0$ is the size of the box), 
Eq.~(\ref{eq:kolmodecay}) becomes $dE/dt \sim E^{3/2}/L_0$, resulting in the 
self-similar decay \cite{Borue1995,Skrbek2000,Biferale2003}
\begin{equation}
E(t) \sim t^{-2}.
\label{eq:boundecay}
\end{equation}

\subsubsection{Unbounded}
In unbounded flows, a similarity solution of Eq.~(\ref{eq:kolmodecay}) 
requires some knowledge of the behavior of the energy containing scale $L$, 
which is in turn related to the evolution of $E(k)$ for low wave numbers. 
In the case of an initial large scale spectrum $\sim k^4$, the 
quasi-invariance of Loitsyanski's integral $I$ (see 
\cite{Davidson2004,Ishida2006}) leads, on dimensional grounds, to 
$I\sim L^5 E$, and replacing in Eq.~(\ref{eq:kolmodecay}) we get 
Kolmogorov's result 
\cite{kolmo1941}
\begin{equation}
E(t) \sim t^{-10/7}.
\label{eq:noboundecay}
\end{equation}
A different decay law is obtained if an initial $\sim k^2$ 
spectrum is assumed for low wave numbers \cite{Saffman1967,Saffman1967b}. 
In the following, we will consider only the bounded or the $\sim k^4$ 
unbounded cases.

\begin{figure}
\includegraphics[width=8cm,height=6cm]{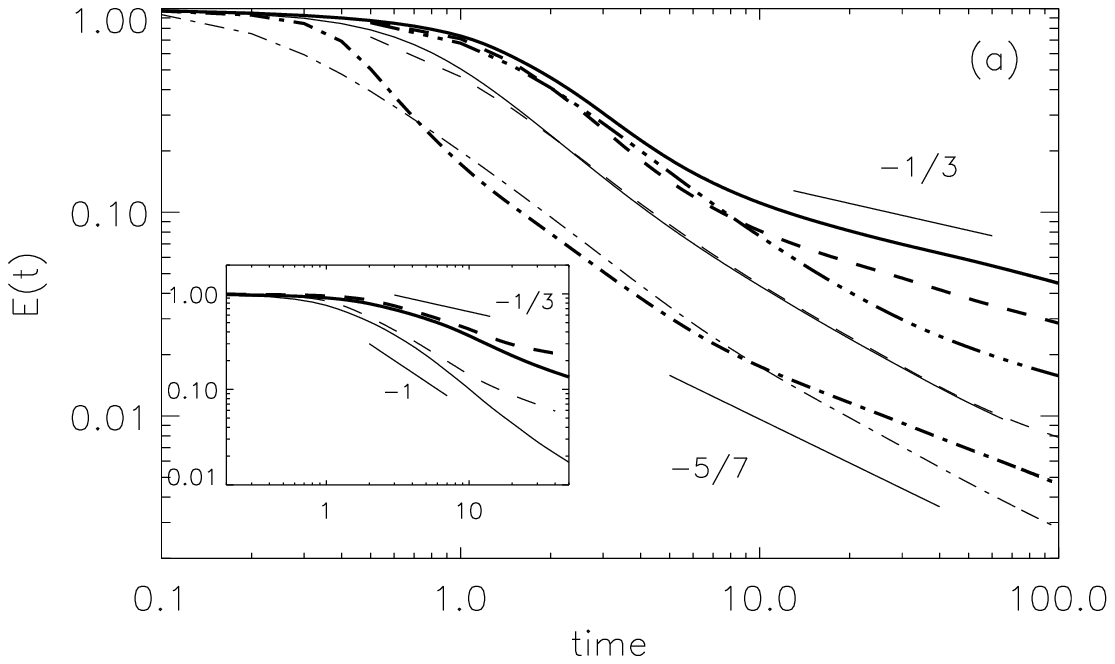}
\includegraphics[width=8cm,height=6cm]{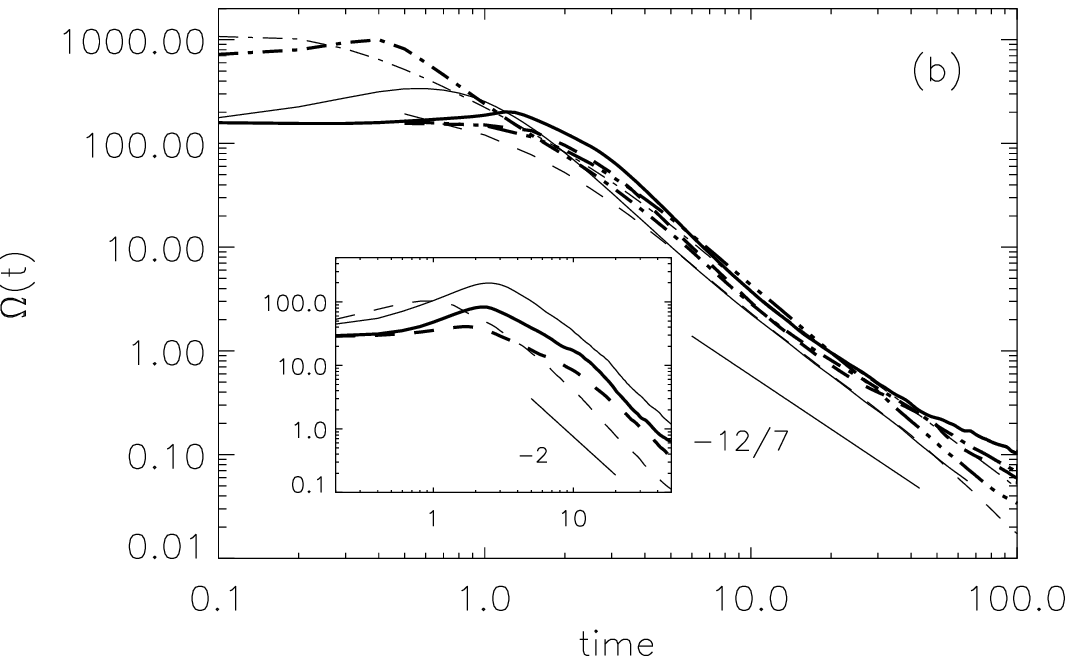}
\includegraphics[width=8cm,height=6cm]{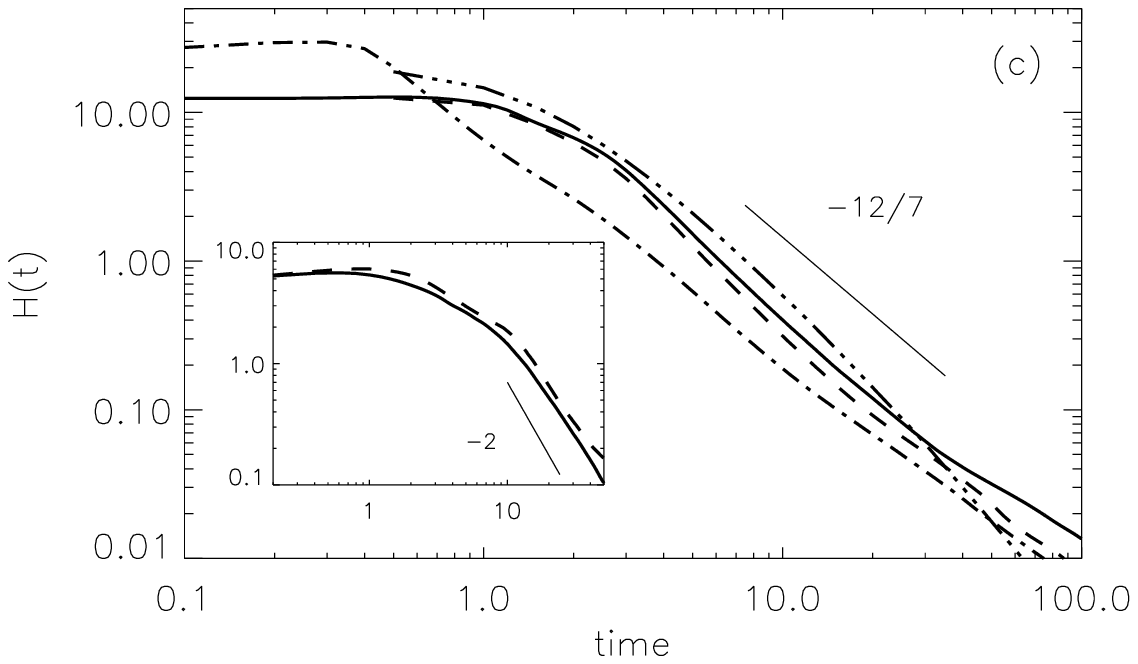}
\caption{(a) Energy decay for rotating unbounded runs ($\Omega=10$). 
Non-helical runs D512-3 (solid), L96-6 (dashed), L192-2 (dash-dotted), and 
helical runs D512H-3 (solid, thick), L96H-3 (dashed, thick), L96H-2 
(dash-tripe-dotted, thick), and L192H-2 (dash-dotted, thick) are 
shown. At late times, the non-helical runs decay slightly faster than 
$t^{-5/7}$, while the helical runs are close to a $-1/3$ decay. The inset 
shows bounded non-helical runs D256-2 (solid) and D512-1 (dashed), and 
helical runs D256H-2 (solid, thick) and D512H-1 (dashed, thick). 
(b) Enstrophy decay for the same runs, with a $-12/7$ slope shown as a 
reference. The inset shows the enstrophy decay in the bounded runs. 
(c) Helicity decay in the unbounded helical runs; bounded helical runs are 
shown in the inset.}
\label{fig:decaysrot}
\end{figure}

\subsection{\label{sec:phenoiso}Rotating flows: isotropic arguments}
\subsubsection{Bounded}
In the case of solid-body rotation without net helicity, a spectra 
$E(k) \sim \epsilon^{1/2}\Omega^{1/2}k^{-2}$ is often assumed at 
small scales (i.e., wave numbers larger than the integral wave number) 
\cite{Zeman1994,Zhou1995,Muller2007,Bellet2006,Mininni2009b}. Replacing 
in the balance equation, this spectrum leads to
\begin{equation}
\frac{dE}{dt} \sim \frac{E^2}{L^2 \Omega}.
\label{eq:rotbalance1}
\end{equation}
For bounded flows $L\sim L_o$ and we get 
\cite{Davidson2006,Morize2006,Bellet2006}
\begin{equation}
E(t) \sim t^{-1}.
\label{eq:boundrot}
\end{equation}

In helical rotating flows the small-scale energy spectrum takes a 
different form. The direct transfer is dominated by the helicity 
cascade. In this case we can write the helicity flux as 
$\delta \sim h_l/(\Omega \tau_l^2)$  where $h_l$ is the helicity at the 
scale $l$, and $\tau_l$ the eddy turnover time \cite{Mininni09}. Constancy 
of $\delta$ leads to small scale spectra $E(k) \sim k^{-n}$ and 
$H(k) \sim k^{n-4}$, where $n=5/2$ obtains for the case of maximum 
helicity \cite{Mininni09}. Further use of dimensional analysis leads to 
$E(k) \sim \epsilon^{1/4}\Omega^{5/4}k^{-5/2}$ for the energy spectrum in 
terms of the energy dissipation rate, and replacing in the balance 
equation we get
\begin{equation}
\frac{dE}{dt} \sim \frac{E^4}{L^{6}\Omega^5}.
\label{eq:rotbalance2}
\end{equation}
For $L\sim L_o$ then \cite{Teitelbaum2009}
\begin{equation}
E(t)\sim t^{-1/3}.
\label{eq:bounrothelidecay}
\end{equation}

\subsubsection{Unbounded}
For non-helical unbounded flows with $E(k)\sim k^4$ at small wave numbers 
we can again make use of the constancy of $I$ in Eq.~(\ref{eq:rotbalance1}), 
leading to \cite{Squires1994}
\begin{equation}
E(t)\sim t^{-5/7}.
\label{eq:nobounrotdecay} 
\end{equation}

For helical flows, assuming $I$ remains constant in 
Eq.~(\ref{eq:rotbalance2}), we obtain \cite{Scripta}
\begin{equation}
E(t)\sim t^{-5/21}.
\label{eq:bounrothelidecay2}
\end{equation}

\begin{figure}
\includegraphics[width=8cm,height=5cm]{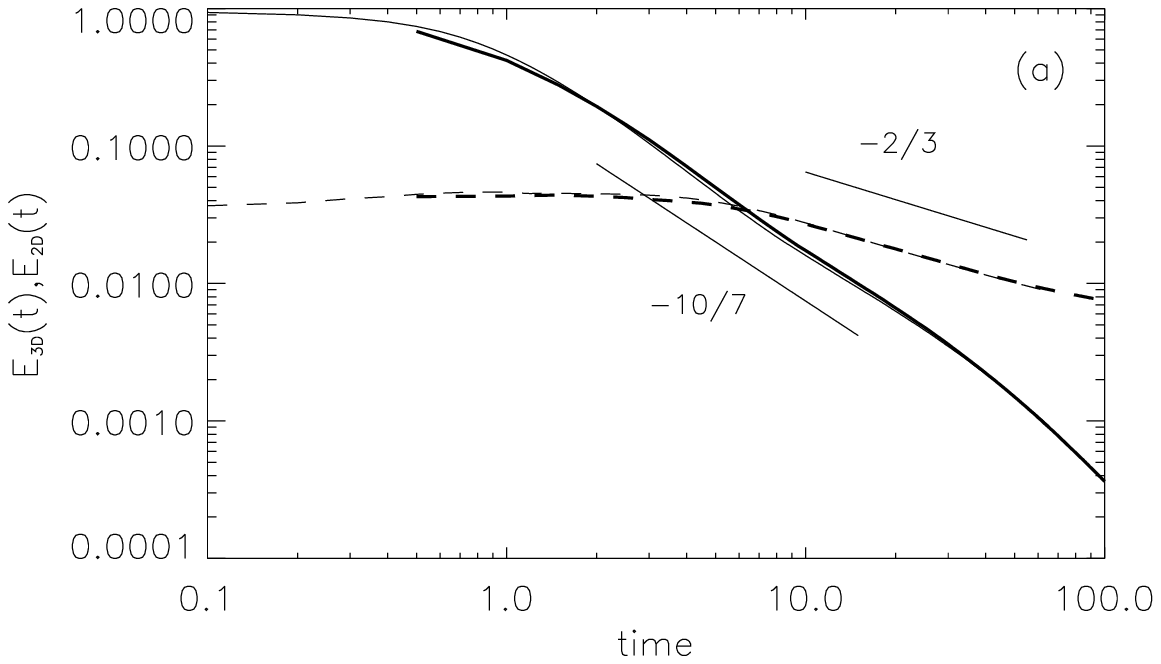}
\includegraphics[width=8cm,height=5cm]{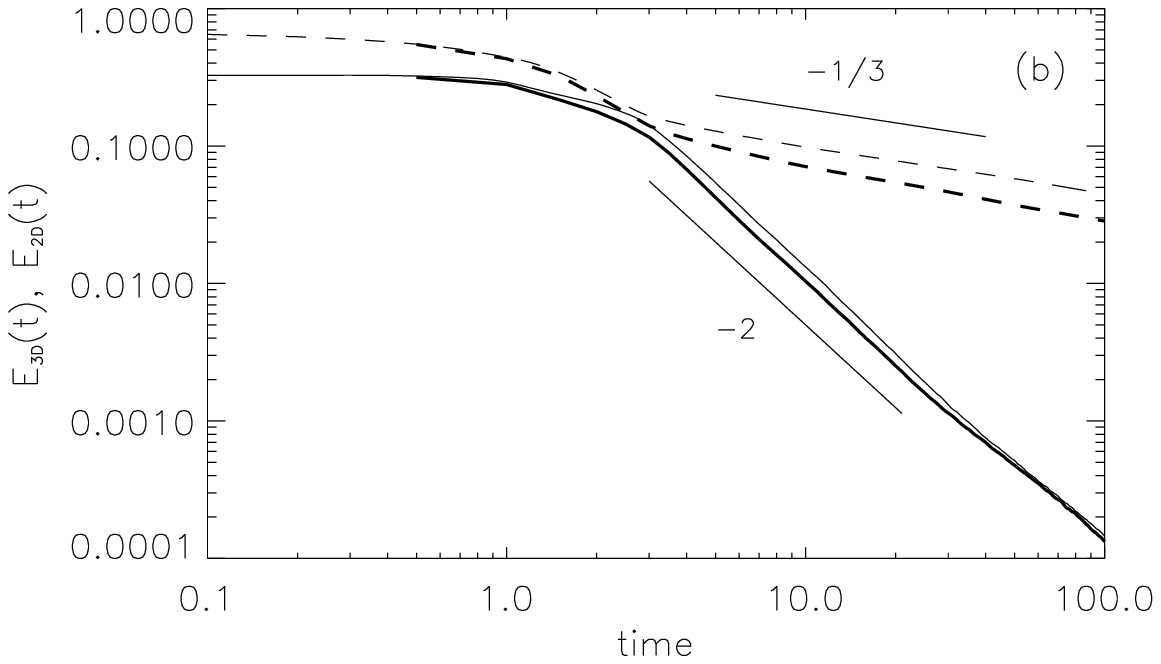}
\includegraphics[width=8cm,height=5cm]{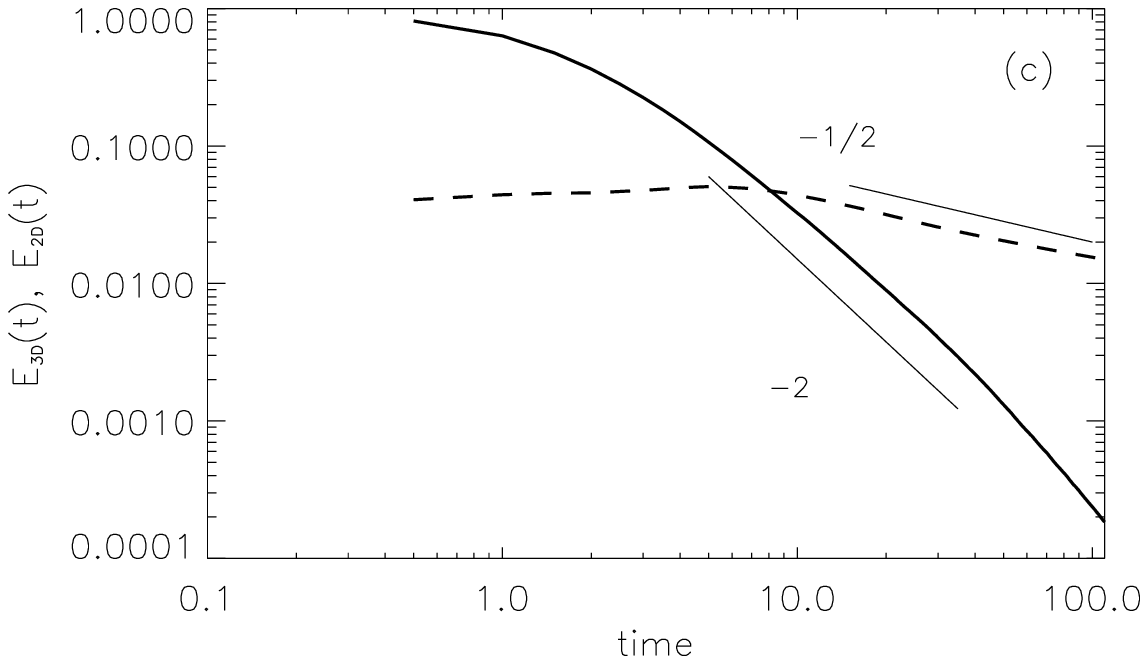}
\includegraphics[width=7.4cm,height=4.6cm]{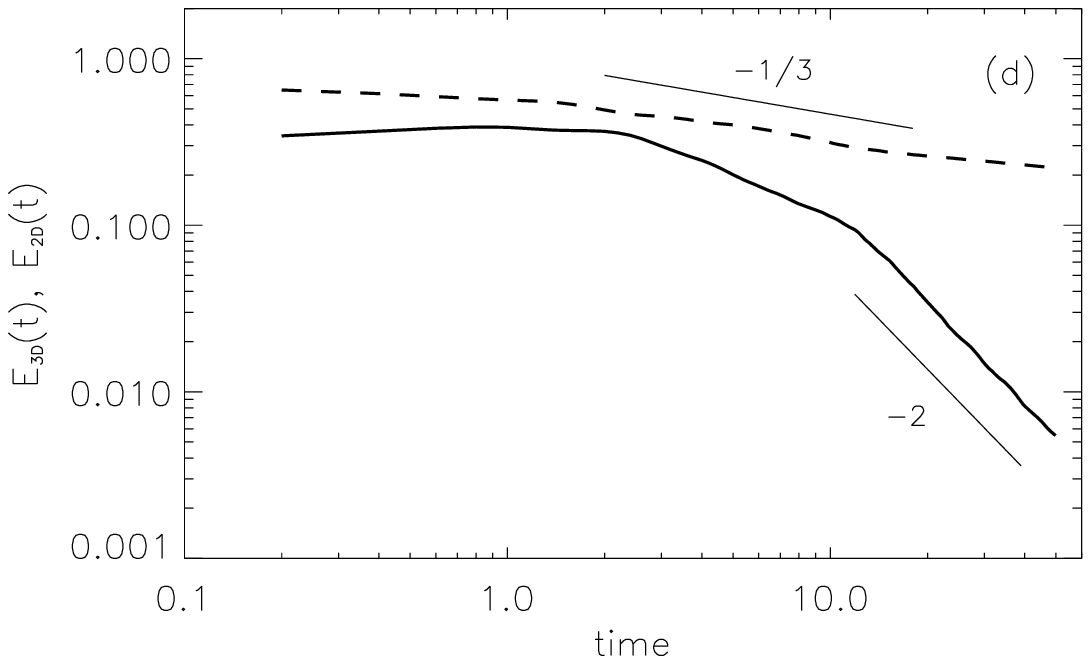}
\caption{Energy decay for $E_{3D}$ (solid) and $E_{2D}$ (dashed) for runs 
with rotation: (a) Unbounded non-helical D512-3 (thin) and L96-6 (thick); 
$E_{3D} \sim t^{-10/7}$ and $E_{2D} \sim t^{-2/3}$ decays are indicated. 
(b) Unbounded helical with random initial conditions L96H-2; $E_{2D}$ is 
close to $t^{-1/2}$. (c)  Unbounded helical with ABC initial conditions 
D512H-3 (thin) and L96H-3 (thick); $E_{2D}$ is close to $t^{-1/3}$. (d) 
Bounded helical with ABC initial conditions D512H-1.}
\label{fig:anidecay1}
\end{figure}

\subsection{Rotating flows: anisotropic arguments}
The decay laws obtained for rotating flows in Eqs.~(\ref{eq:boundrot}), 
(\ref{eq:bounrothelidecay}), and (\ref{eq:nobounrotdecay}), have been 
reported in experiments or in simulations 
\cite{Yang2004,Morize2005,Morize2006,Teitelbaum2009}. However, the 
analysis above is based on the isotropic energy spectrum and on 
the quasi-invariance of the isotropic Loitsyanski integral. For an 
anisotropic flow, other quantities are expected to be 
quasi-invariants during the decay instead 
\cite{Batchelor,Davidson2004,Davidson2010}.

Rotating flows tend to become quasi-2D, and the assumption of an 
axisymmetric energy spectrum seems natural considering the symmetries 
of the problem. If there is no dependence on wave numbers on the 
parallel direction, the energy spectrum for small values of $k_\perp$ 
can be expanded as (see, e.g., \cite{Fox2008})
\begin{equation}
E(k_\perp) \approx L k_\perp^{-1} + K k_\perp + I_{2D} k_\perp^3 + \cdots
\end{equation}
We will be interested in the following coefficients:
\begin{equation}
K=\int{\langle {\bf u \cdot u'} \rangle r \, dr} ,
\end{equation}
and 
\begin{equation}
I_{2D}=\int{\langle {\bf u \cdot u'} \rangle \, r^3 dr} ,
\end{equation}
where $\langle {\bf u \cdot u'} \rangle$ is the two-point correlation 
function for spatial increments $r$ perpendicular to the rotation 
axis. If the correlation function decays fast enough for large 
values of $r$ \cite{Batchelor}, these quantities can be expected to be 
quasi-invariants during the decay, respectively for initial large-scale 
energy spectra $\sim k_\perp$ and $\sim k_\perp^3$, in the same way $I$ 
is quasi-conserved during the decay of isotropic flows with an initial 
large-scale $\sim k^4$ energy spectrum. A detailed proof of the 
conservation of $K$ for rotating flows can be found in \cite{Davidson2010}; 
it is a direct consequence of the conservation of linear momentum in the 
direction parallel to the rotation axis. It is worth pointing out that 
these quantities were also shown to be conserved in other systems: 
proofs of the conservation of $K$ and $I_{2D}$ for quasigeostrophic 
flows can be found in \cite{Fox2008}. In practice, these quantities are 
only approximately conserved in numerical simulations, see e.g., the 
approximate constancy of $I_{2D}$ and $K$ reported for rotating flows 
in \cite{Scripta}.

As per virtue of the decay the Rossby number decreases with time, 
we will further assume for our phenomenological analysis that 2D and 3D 
modes are only weakly coupled, and write equations for the energy in the 
2D modes, $E_{2D}$. In the non-helical case, if $K$ remains approximately 
constant with $K \sim E_{2D} L_{\perp}^2 L_{0\parallel}$ (where 
$L_{0\parallel}$ is the size of the box in the direction parallel to 
$\mbox{\boldmath $\Omega$}$), then Eq.~(\ref{eq:rotbalance1}) for the 2D 
modes becomes
\begin{equation}
\frac{dE_{2D}}{dt} \sim \frac{E_{2D}^3L_{0\parallel}}{K \Omega},
\end{equation}
which leads to a decay
\begin{equation}
E_{2D}(t)\sim t^{-1/2}.
\label{eq:bounrotdecay2D}
\end{equation}
Alternatively, constancy of $I_{2D} \sim E_{2D} L_{\perp}^4 L_{0\parallel}$ 
in  Eq.~(\ref{eq:rotbalance1}) for the 2D modes leads to
\begin{equation}
\frac{dE_{2D}}{dt} \sim \frac{E_{2D}^{5/2}L_{0\parallel}^{1/2}}
    {I_{2D}^{1/2}\Omega} ,
\end{equation}
and
\begin{equation}
E_{2D}(t)\sim t^{-2/3}.
\label{eq:bounrotdecay2D2} 
\end{equation}

\begin{figure}
\begin{center}
\includegraphics[width=8cm,height=6cm]{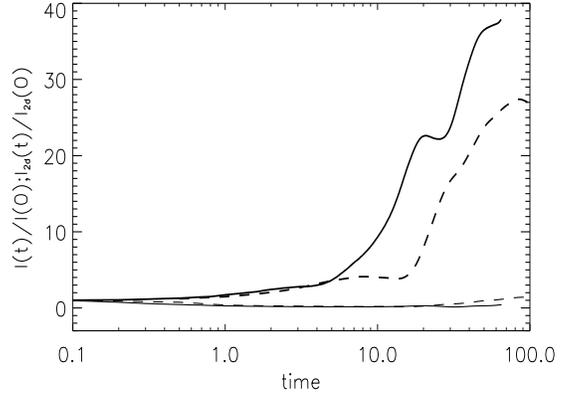}
\caption{Evolution of $I/I(0)$ (thick lines) and of $I_{2D}/I_{2D}(0)$ 
(thin lines) for runs D512-3 (solid) and L96-6 (dashed). While in both runs 
$I_{2D}$ mantains an approximately constant value, $I$ growths monotonically 
and during the self-similar energy decay increases by one order of 
magnitude.}
\label{fig:Loit}
\end{center}
\end{figure}

The same arguments can be extended to the helical rotating case using 
Eq.~(\ref{eq:rotbalance2}). If constancy of $K$ is assumed we get
\begin{equation}
\frac{dE_{2D}}{dt} \sim \frac{E_{2D}^7 L_{0\parallel}^3}{\Omega^5 K^3} ,
\end{equation}
and
\begin{equation}
E(t)\sim t^{-1/6} .
\label{eq:bounrothelidecay2D} 
\end{equation}
Finally, constancy of $I_{2D}$ leads to
\begin{equation} 
\frac{dE_{2D}}{dt} \sim 
    \frac{E_{2D}^{11/2}L_{0\parallel}^{3/2}}{I_{2D}^{3/2}\Omega^5} ,
\end{equation}
and
\begin{equation}
E(t)\sim t^{-2/9}.
\label{eq:bounrothelidecay2D2}
\end{equation}
These decay laws will be important to analyze the evolution of the energy 
in the simulations discussed in the next section.

\subsection{Enstrophy decay}
From any of the previous energy decay laws, one can also compute laws for 
the enstrophy decay $\Omega(t) = \left< \omega^2 \right>/2$ using the 
isotropic energy balance equation and replacing $\epsilon = \nu \Omega(t)$, 
which results in $\Omega(t) = \nu^{-1} dE/dt$. From this equation, for 
every solution for which the energy decays as $E(t) \sim t^{\alpha}$, the 
enstrophy decay results
\begin{equation}
\Omega(t)\sim t^{\alpha -1}.
\label{eq:decayenstro} 
\end{equation}

Although rotating flows are anisotropic, the enstrophy is predominantly a 
small-scale magnitude and we will see that this isotropic argument gives 
good agreement with the numerical results for rotating and non-rotating 
flows. Since helicity is related with the energy and the enstrophy only 
through a Schwartz inequality, no simple decay laws can be derived in its 
case using these phenomenological arguments.

\section{\label{sec:decay}Time evolution - Numerical Results}

We present here the results for the energy, enstrophy, and helicity decay 
obtained in the numerical simulations listed in Table \ref{tab:paratable}, 
classifying them as rotating or non-rotating, bounded or unbounded (in 
the sense that the initial integral scale is smaller than the size of the 
box), and helical or non-helical.

Concerning the terminology of ``bounded'' and ``unbounded'' used to 
describe the numerical simulations, it is important to note that 
confinement effects in a rotating flow go beyond a saturation of the 
integral scale when it grows to the box size. Confinement also selects 
a discrete set of inertial waves which are normal modes of the domain, 
and boundaries can introduce dissipation through Ekman layers. The latter 
effect is not present in our numerical simulations with periodic boundary 
conditions. Finally, it was shown in \cite{Wang2002} (see also 
\cite{Pope2000}) that the small number of Fourier modes available in the 
shells with wave number $k \approx 1$ gives rise to poor representation 
of isotropy and of the integral scale in runs for which the integral 
scale approaches $1/5$ of the box size. As a result, the ``bounded'' runs 
are here only briefly considered to study the time evolution of global 
quantities (energy, enstrophy, and helicity), and to compare with the 
prediction obtained in the corresponding cases in the phenomenological 
analysis.

\subsection{Non-rotating flows}
Numerical results for non-rotating, bounded and unbounded flows are 
shown in Fig.~\ref{fig:decays}. In the unbounded case (runs with an 
initial energy spectrum $\sim k^4$ peaking at $k=14$ in the DNS and $96^3$ 
LES, and peaking at $k=30$ in the $192^3$ LES), the runs show a decay for 
the energy close to $\sim t^{-10/7}$ independently of the presence of 
helicity or not (note the runs also span a range of Reynolds numbers from 
$Re \approx 420$ to $1200$). The decay is consistent with the prediction 
given by Eq.~(\ref{eq:noboundecay}) for an initial $\sim k^4$ energy 
spectrum. 

\begin{figure}
\includegraphics[width=8cm,height=6cm]{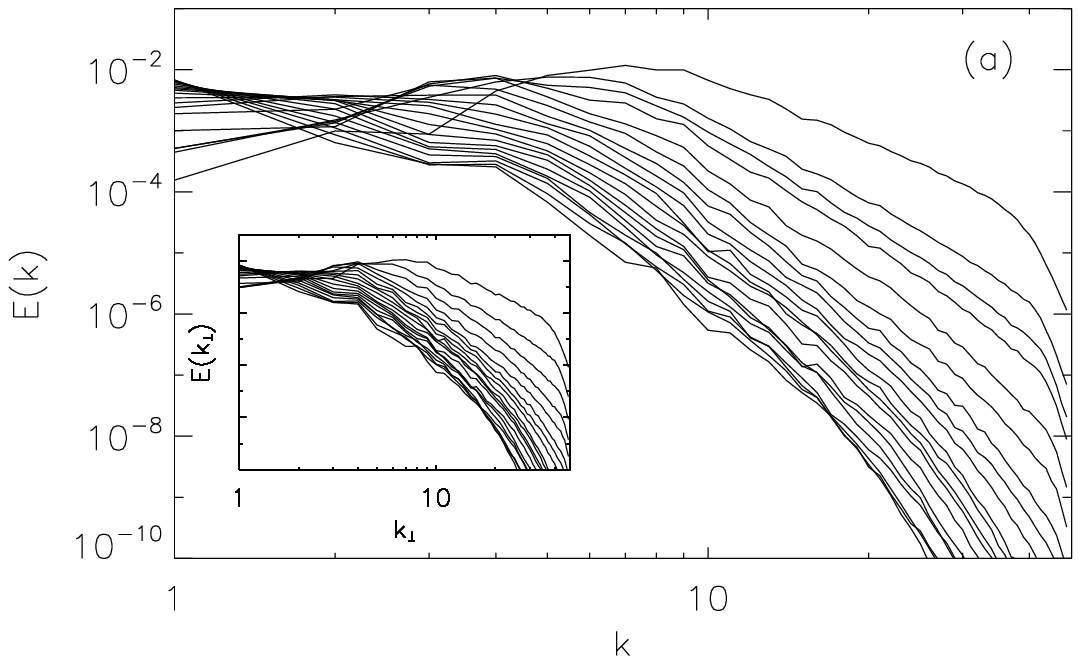}
\includegraphics[width=8cm,height=6cm]{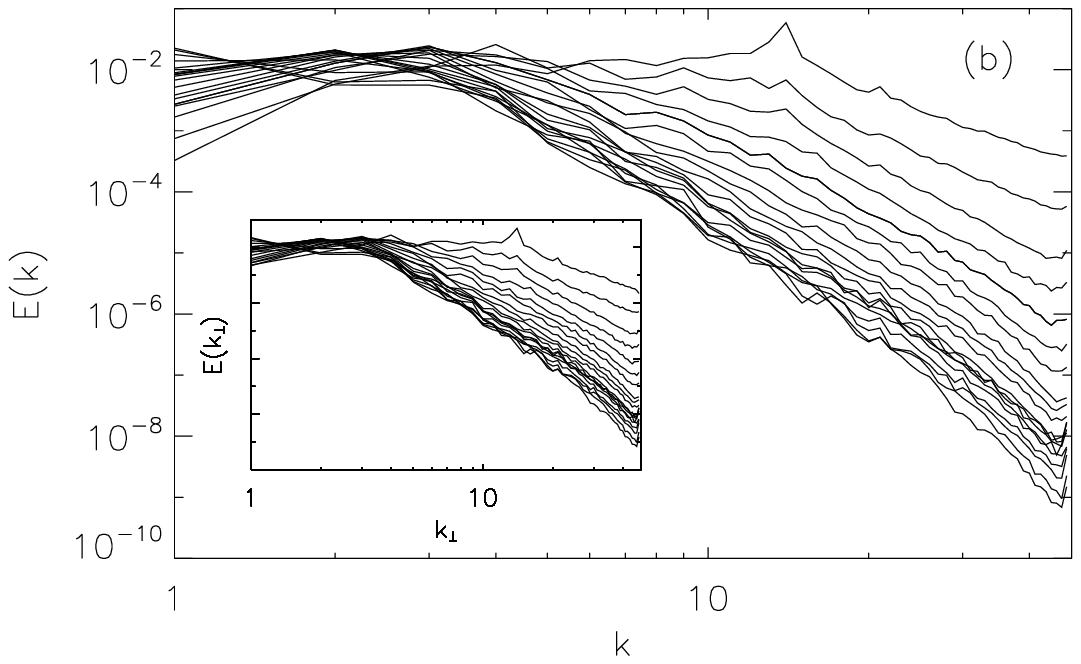}
\caption{(a) Evolution of the isotropic energy spectrum $E(k)$ for L96-6 
(non-helical, $\Omega=10$, initial $\sim k^4$ spectrum peaking at $k=14$) 
from $t=5$ to $t=100$ with time increments $\Delta t=5$. Inset: reduced 
perpendicular energy spectrum $E(k_\perp)$ for the same times. 
(b) Evolution of the isotropic energy spectrum for L96H-2 (helical, 
$\Omega=10$, initial $\sim k^4$ spectrum peaking at $k=14$) at the same 
times, with the reduced perpendicular energy spectrum in the inset.}
\label{fig:espectros}
\end{figure}


The enstrophy decay is also consistent with this law, as expressed by 
Eq.~(\ref{eq:decayenstro}), decaying close to $\sim t^{-17/7}$ in all 
cases. In the absence of rotation, helicity only delays the onset of the 
self-similar decay by retarding the time when the maximum of enstrophy 
takes place, as already reported in \cite{Morinishi2001} and 
\cite{Lesieur1977}. This is more clearly seen in the DNS; see, e.g., 
the time of the peak of enstrophy for runs D512-3 and D512H-3 in 
Fig.~\ref{fig:decays}(b). Finally, in the helical runs, helicity seems to 
decay as the enstrophy, just slightly slower than the $\sim t^{-17/7}$ 
law. 

Similar results are observed for bounded flows, i.e., for initial 
conditions with the initial energy containing scale close to the size of 
the box (runs with a $\sim k^{-4}$ spectrum from $k=4$ to $14$, peaking 
at $k=4$). In this case, all runs are consistent with a $\sim t^{-2}$ 
decay for the energy (see the insets of  Fig.~\ref{fig:decays}) in 
agreement with Eq.~(\ref{eq:boundecay}), and a decay for the enstrophy 
close to $\sim t^{-3}$ in agreement with Eq.~(\ref{eq:decayenstro}). In 
the helical runs, helicity decays again slightly slower than the enstrophy, 
but close to the $\sim t^{-3}$ power law.

\subsection{Rotating flows}

\subsubsection{Global quantities}
As rotation is increased, the simulations show a shallower power law in 
the energy decay. As an illustration, Fig. \ref{fig:decay_dif_rot_rates} 
shows the energy decay rate in simulations of unbounded non-helical flows 
with increasing rotation rate $\Omega$. As reported in previous numerical 
simulations \cite{Yang2004} and experiments 
\cite{Morize2005,Morize2006}, as $\Omega$ increases the decay 
slows down until reaching a saturated decay for $Ro \approx 0.1$. We will 
focus in the following in simulations with Rossby number small enough to 
observe this saturated decay, although not so small that the rotation 
quenches all non-linear interactions giving only exponential decay. A 
detailed study of the transition between the non-rotating and rotating 
cases can be found in \cite{Morize2005}.

Figure \ref{fig:decaysrot} shows the energy, enstrophy, and helicity 
decay in simulations of rotating flows with and without helicity, in the 
unbounded and bounded cases (the latter in the insets). The energy decay 
in the unbounded non-helical runs (thin lines in Fig.~\ref{fig:decaysrot}) 
is slightly steeper than what Eq.~(\ref{eq:nobounrotdecay}) predicts 
($E \sim t^{-5/7}$). A better agreement is observed for the enstrophy, 
which is closer to a $\sim t^{-12/7}$ law. As will be shown next, the 
agreement between the phenomenological arguments for the energy and the 
simulations is improved if the decay of 2D and 3D modes is considered 
separately.

\begin{figure}
\begin{center}
\includegraphics[width=9cm,height=8.5cm]{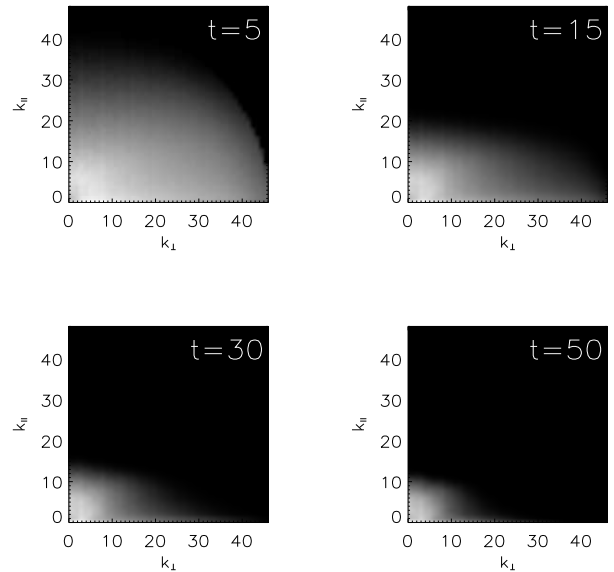}
\caption{Axisymmetric energy spectrum $e(k_\parallel,k_\perp)/\sin\theta$ 
for different times for run L96-6 (non-helical, $\Omega=10$, initial 
energy spectrum $\sim k^4$ peaking at $k=14$).}
\label{fig:contour2}
\end{center}
\end{figure}

Alternatively, the unbounded helical runs (thick lines in 
Fig.~\ref{fig:decaysrot}) show for the energy a $\sim t^{-1/3}$ decay or 
steeper (although shallower than $\sim t^{-5/7}$). Runs with ABC initial 
conditions tend to develop a clearer power law decay and to be closer to 
a $\sim t^{-1/3}$ decay than runs with random helical modes. Again, 
these differences can be explained considering the decay of 2D and 
3D modes, as well as the effect of anisotropy in the initial conditions 
which is specially relevant for this particular case. The enstrophy and 
helicity show a decay close to $\sim t^{-12/7}$. Note that in the presence 
of rotation, helicity not only slows down the occurrence of the peak of 
enstrophy as already reported in \cite{Morinishi2001}, but it also changes 
the energy decay after this peak. The enstrophy decay is not affected by 
the presence of helicity.

Overall, the case of constrained runs shows a similar scenario, with a 
significant slow down of the decay rates in the presence of rotation, and 
with an extra slow down of energy decay in the presence of helicity 
(see the insets in Fig.~\ref{fig:decaysrot}). Rotating non-helical flows 
are close to $E(t) \sim t^{-1}$, $\Omega(t) \sim t^{-2}$, and 
$H(t) \sim t^{-2}$, while helical flows in this case display a shallower 
decay in the energy consistent with $E \sim t^{-1/3}$ as predicted by the 
phenomenological arguments that take into account the effect of helicity 
in the energy spectrum of rotating turbulence. As in the unbounded case, 
the presence of helicity does not affect the decay rate of enstrophy. 

\subsubsection{Anisotropic global quantities}
Although the impact of rotation and helicity in the energy decay is clear, 
the predictions given by the isotropic phenomenological arguments in 
Sec.~\ref{sec:phenoiso} do not coincide in all cases with the results 
from the simulations. This can be ascribed to the fact that these 
arguments assume an isotropic energy scaling, while rotation breaks down 
isotropy making spectral energy distribution become axisymmetric with 
the preferred direction along the axis of rotation. Two-dimensionalization 
of the flow has already been reported during the decay 
\cite{Morinishi2001,Yang2004,Kuczaj}, as well as weak coupling of 
2D and 3D modes \cite{Chen2005,Bourouiba2008} for very small Rossby numbers. 
Based on this, we discriminate between the energy contained in 3D modes with 
$k_z \neq 0$ ($E_{3D}$), and the energy in slow 2D modes with 
$k_z=0$ ($E_{2D}$). At this point it is important to point out that 
as the energy decays in the simulations, the Rossby number also 
monotonically decreases with time. As a rule, the Rossby numbers decrease 
by one order of magnitude in the first turnover times (from the values 
listed in Table \ref{table:runs}, which correspond to the time of maximum 
dissipation), and another order of magnitude at $t\approx 100$.

\begin{figure}
\begin{center}
\includegraphics[width=9cm,height=8.5cm]{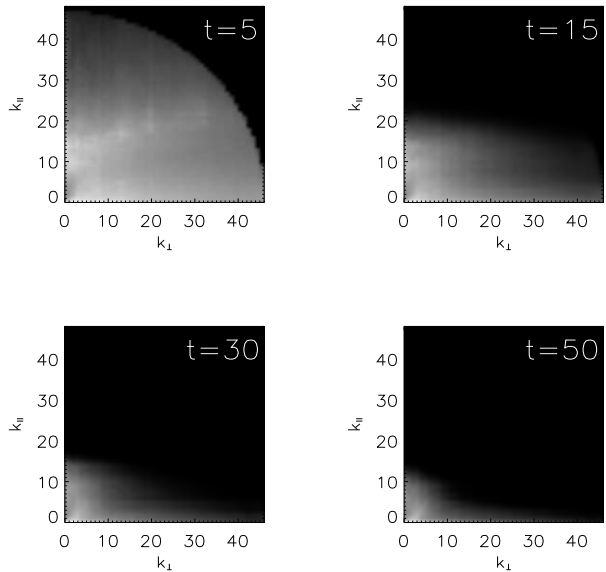}
\caption{Axisymmetric energy spectrum $e(k_\parallel,k_\perp)/\sin\theta$ 
for different times for run L96H-3 (helical with ABC initial conditions, 
$\Omega=10$, initial energy spectrum $\sim k^4$ peaking at $k=14$).}
\label{fig:contour3}
\end{center}
\end{figure}

In Fig.~\ref{fig:anidecay1} we show $E_{3D}$ and $E_{2D}$ as a function 
of time for several runs. In each and every case we can clearly identify 
distinct behaviors for the 2D and 3D energies, obeying different power-law 
decays. On the one hand, $E_{3D}$ always shows a decay close to some power 
law expected for a (bounded or unbounded) non-rotating case, with the 
unbounded non-helical runs having $E_{3D} \sim t^{-10/7}$ in agreement 
with Eq.~(\ref{eq:noboundecay}) as illustrated in 
Fig.~\ref{fig:anidecay1}(a), and with most helical runs (bounded and 
unbounded) with $E_{3D} \sim t^{-2}$ in agreement with 
Eq.~(\ref{eq:boundecay}) (which corresponds to the bounded non-rotating 
decay) as illustrated in Fig.~\ref{fig:anidecay1}(b--d). On the other hand, 
$E_{2D}$ follows power laws close to the ones predicted by 
Eqs.~(\ref{eq:bounrotdecay2D})--(\ref{eq:bounrothelidecay2D2}). The 
unbounded non-helical runs are compatible with $E_{2D} \sim t^{-2/3}$, 
and the helical runs show $\sim t^{-1/2}$ or $\sim t^{-1/3}$ (note in the 
helical case the power laws predicted are for the case of maximum helicity, 
and for intermediate helicity the power laws are bounded between the 
non-helical and the maximally helical values).

The results in Fig.~\ref{fig:anidecay1} indicate clearer power law 
decay (and better agreement with phenomenological arguments) is obtained for 
the separate energy in 2D and 3D modes, than when the total energy is 
considered (compare, e.g., the extent of the power laws in this 
figure with the ones in Fig.~\ref{fig:decaysrot}). This is clearer in the 
non-helical case, where all unbounded runs show a decay consistent with 
Eq.~(\ref{eq:bounrotdecay2D2}) for the 2D modes (all non-helical runs with 
random initial conditions have a $\sim k_\perp^3$ initial energy spectrum, 
per virtue of the isotropic initial $\sim k^4$ spectrum), and with 
Eq.~(\ref{eq:noboundecay}) for the 3D modes. The separate evolution of 
$E_{3D}$ and $E_{2D}$ seems to be independent of the initial ratio of energy 
in 3D and 2D modes, at least for the range of parameters explored in this 
work.

To further show the agreement with the phenomenological arguments, 
the evolution of $I$, $I_{2D}$ and $K$ must be considered, to verify whether 
these quantities behave in agreement with the assumptions used to derive the 
decay laws in Sec.~\ref{sec:phenom}. Figure \ref{fig:Loit} shows the time 
evolution of $I$ and $I_{2D}$ for two simulations with rotation, normalized 
by their initial values at $t=0$. In both cases $I$ grows monotonically in 
time by a factor of $\approx 50$ during the decay. This fast increase of 
$I$ is observed in all rotating runs. Meanwhile, $I_{2D}$ settles and 
remains approximately constant, fluctuating around its initial value.

The helical runs show more disparity in the time evolution of the 2D and 
3D energies. Bounded runs show $E_{2D} \sim t^{-1/3}$ and 
$E_{3D} \sim t^{-2}$, which agree with the previous scenario 
where 3D modes decay as in the non-rotating case, and slow 2D modes 
decay according to the anisotropic prediction with rotation (in this 
case, corresponding to Eqs. (\ref{eq:bounrothelidecay}) and 
(\ref{eq:boundecay}), respectively). But for initial conditions that peak 
at $k \approx 14$, in some cases they show decays of $E_{3D}$ and $E_{2D}$ 
that are consistent with predictions for bounded flows 
(Fig.~\ref{fig:anidecay1} (b)), while in others they show decays as in the 
unbounded case (Fig.~\ref{fig:anidecay1} (c)).
It may be the case that in the presence of helicity more separation of 
scales is needed between the initial energy containing scale and the 
size of the box in order to study unbounded flows (indeed, run L192H-2, 
which has an initial energy spectrum peaking at $k=30$, shows a decay 
compatible with $E_{3D} \sim t^{-10/7}$). But it is also observed that 
these decay laws also depend on whether ABC or random helical initial 
conditions are used. In the ABC flow, two-thirds of the initially 
excited modes are in the $k_{\parallel}=0$ plane (see Eq.~(\ref{eq:ABC})), 
while random initial conditions excite modes in Fourier space distributed 
more isotropically, resulting in a smaller percentage of energy in the 
$k_{\parallel}=0$ modes when compared with the energy in 
$k_{\parallel} \neq 0$. This dependence in the initial ratio of energy 
in 2D and 3D modes may indicate a stronger coupling between 2D 
and 3D modes in the presence of helicity (in Sec.~\ref{sec:anisotropy} 
we will explicitly show how changes in the initial anisotropy affect 
these results).

\begin{figure}
\includegraphics[width=8cm,height=6cm]{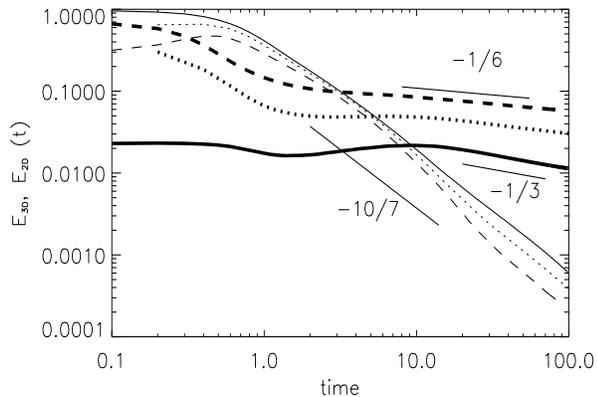}
\caption{Energy decay for different initial anisotropies. Thick lines 
correspond to the energy in $2D$ modes ($E_{2D}$), while thin lines 
correspond to energy in $3D$ modes ($E_{3D}$). Simulations shown are 
L192HA-1 (solid), L192HA-2 (dotted), and L192HA-3 (dashed), with 
increasing anisotropy.}
\label{fig:decayani}
\end{figure}


On dimensional grounds, the impact of helicity in the coupling can be 
explained as follows. If decoupling takes place in the limit of fast 
rotation, it should hold until a time $t^* \sim Ro^{-2}$, after which 
higher order terms in the multiple time scale expansion make non-resonant 
interactions relevant \cite{Babin1996,Bourouiba2008}. In non-helical 
unbounded turbulence, $E\sim t^{-5/7}$ and $L\sim t^{1/7}$ (under approximate 
conservation of $I$). The Rossby number then decays as
\begin{equation}
Ro = \frac{E^{1/2}}{2^{1/2} L \Omega} \sim t^{-1/2} ,
\end{equation}
and $t^*$ grows as $t$. As a result, if decoupling takes place in the 
freely decaying non-helical case, it can be sustained for long times. 
The same result ($Ro \sim t^{-1/2}$) is obtained if the argument is 
refined to consider the 2D invariants $K$ or $I_{2D}$ using 
Eqs.~(\ref{eq:bounrotdecay2D}) or (\ref{eq:bounrotdecay2D2}), or in the 
bounded case using Eq.~(\ref{eq:boundrot}). However, in the helical case 
(e.g., using Eq.~(\ref{eq:bounrothelidecay})) a much slower decay of the 
Rossby number obtains
\begin{equation}
Ro \sim t^{-1/6},
\end{equation}
and thus $t^*$ grows only as $t^{1/3}$.

\subsection{\label{subsec:espectros}Spectral evolution and anisotropy}

\begin{figure}
\includegraphics[width=8cm,height=6cm]{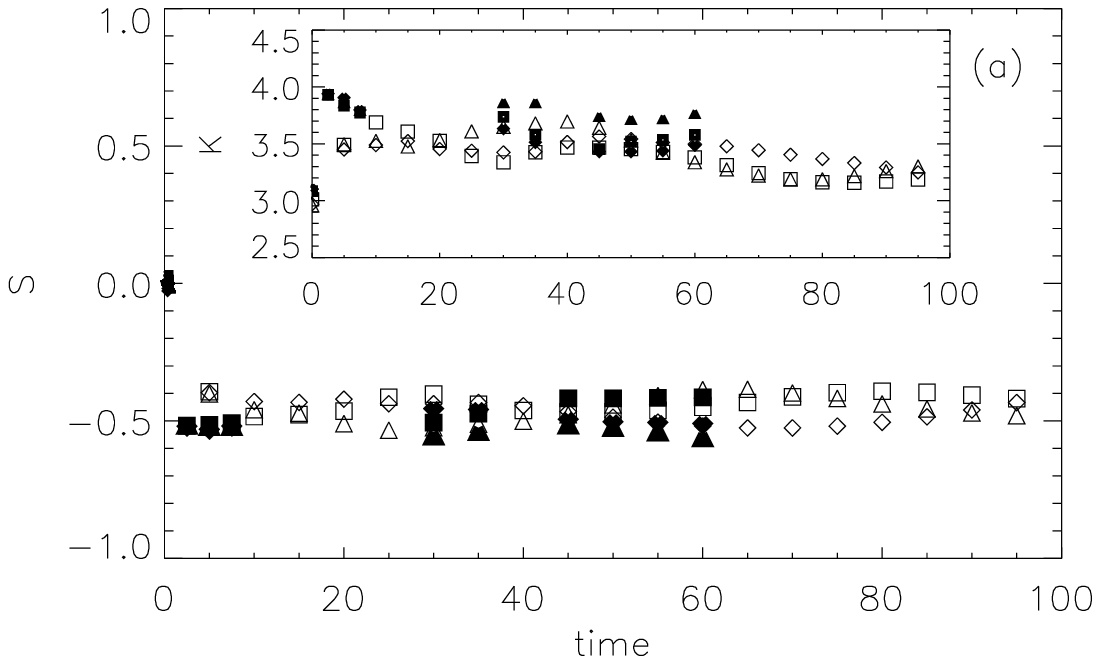}
\includegraphics[width=8cm,height=6cm]{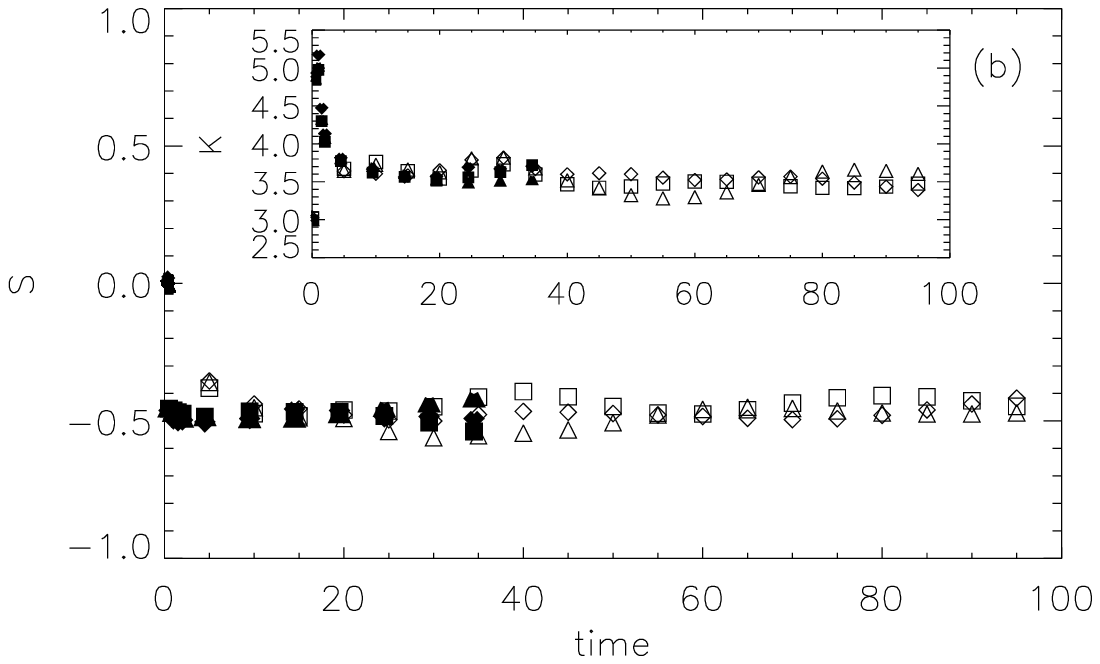}
\caption{Evolution of the velocity-derivative skewness for non-rotating 
runs (a) D512-2 and L96-1 (non-helical), and (b) D512H-2 and L96H-1 
(helical). DNS have filled symbols while LES have empty symbols, with 
squares for $S_x$, triangles for $S_y$, and diamonds for $S_z$. The 
three components of $S$ oscillate around $\approx -0.5$ independently 
of helicity content. The insets show the three components of the 
kurtosis for the same runs using the same labels.}
\label{fig:skewness}
\end{figure}

The isotropic and reduced perpendicular energy spectra are shown in 
Fig.~\ref{fig:espectros} for LES of rotating flows ($\Omega=10$) with 
and without helicity. Energy at large scales grows in all cases, 
indicating an inverse energy transfer (as also evidenced by a negative 
flux of energy in that range). Also, the energy spectrum in the helical 
case (e.g., at the time of maximum enstrophy; not shown) is slightly 
steeper than in the non-helical case (see, e.g., 
\cite{Mininni10a,Mininni10b}).

To further investigate the energy spectral distribution among different 
directions, we show in Figs.~\ref{fig:contour2} and \ref{fig:contour3} 
plots of the axisymmetric energy spectrum $e(k_{\parallel},k_{\perp})$ 
for runs L96-6 and L96H-3. Note that to obtain contour levels 
corresponding to circles in the isotropic case, here and in the following, 
the axisymmetric energy spectrum is divided by $\sin \theta$, where 
$\theta=\arctan(k_\parallel/k_\perp)$. 

In the case without rotation the spectrum has an isotropic 
distribution of energy evidenced by circular contour levels, which 
maintain their shape as the flow decays (not shown). Alternatively, when 
rotation is present, the distribution of energy becomes anisotropic with 
more energy near the $k_\parallel=0$ axis at late times 
(Fig.~\ref{fig:contour2}). This preferential transfer towards slow 
2D modes is well known, see e.g., 
\cite{Cambon1989,Waleffe1993,Cambon1997}. However, for helical 
rotating flows there is even more energy near the $k_\parallel=0$ axis 
(Fig.~\ref{fig:contour3}), and energy is also more concentrated at large 
scales (small $k_\perp$ wave numbers), in agreement with our previous 
observations of a faster increase of integral scales in the presence of 
helicity.

\section{\label{sec:anisotropy}Effect of initial anisotropy}

As mentioned before, some of the differences observed in the evolution 
of $E_{2D}$ and $E_{3D}$ in helical runs are associated with 
differences in the initial conditions. In particular, runs with ABC 
and with random helical initial conditions differ in the fact that the 
ABC flow initially has $2/3$ of the excited modes in Fourier space in 
the slow 2D manifold, while in the random case energy is more 
isotropically distributed.

To further investigate this effect, we consider a set of helical runs 
with random initial conditions but with increasing initial anisotropy 
(runs L192HA-1, L192HA-2, and L192HA-3). Anisotropy is introduced 
by weighting the amplitude of all modes with $k_{\parallel}=0$ with a 
parameter $\alpha$ ($\alpha=1$ corresponds to the isotropic initial 
conditions considered before, and $\alpha>1$ corresponds to larger 
amplitude of the $2D$ modes relative to the $3D$ modes). The runs have 
$\alpha=1$ (L192HA-1), 5 (L192HA-2), and 10 (L192HA-3), resulting 
in initial ratios of the energy in 2D to 3D modes 
$E_{2D}/E_{3D} \approx 0.024$, $0.626$, and $2.408$, respectively.
The runs also have initial energy and helicity spectra peaking at $k=25$, 
thus allowing us to study unbounded cases with larger scale separation.

Figure \ref{fig:decayani} shows that $E_{3D}$ decays approximately as 
$\approx -10/7$ regardless of the anisotropy of the initial conditions, 
while $E_{2D}$ changes its decay becoming shallower as anisotropy grows. 
On the one hand, the isotropic case ($\alpha=1$) is closer to a 
$E_{2D} \sim t^{-1/3}$ law, a result consistent with the $2D$ decay shown 
in Fig.~\ref{fig:anidecay1}(d). On the other hand, the decay in the most 
anisotropic case ($\alpha=10$) is closer to $\sim t^{-1/6}$, which is 
consistent with the decay for helical flows in the case when $K$ is 
approximately constant; see Eq.~(\ref{eq:bounrothelidecay2D}). Indeed, 
it was verified that $K$ remains approximately constant during the 
decay of this run (not shown).

\begin{figure}
\includegraphics[width=8cm,height=6cm]{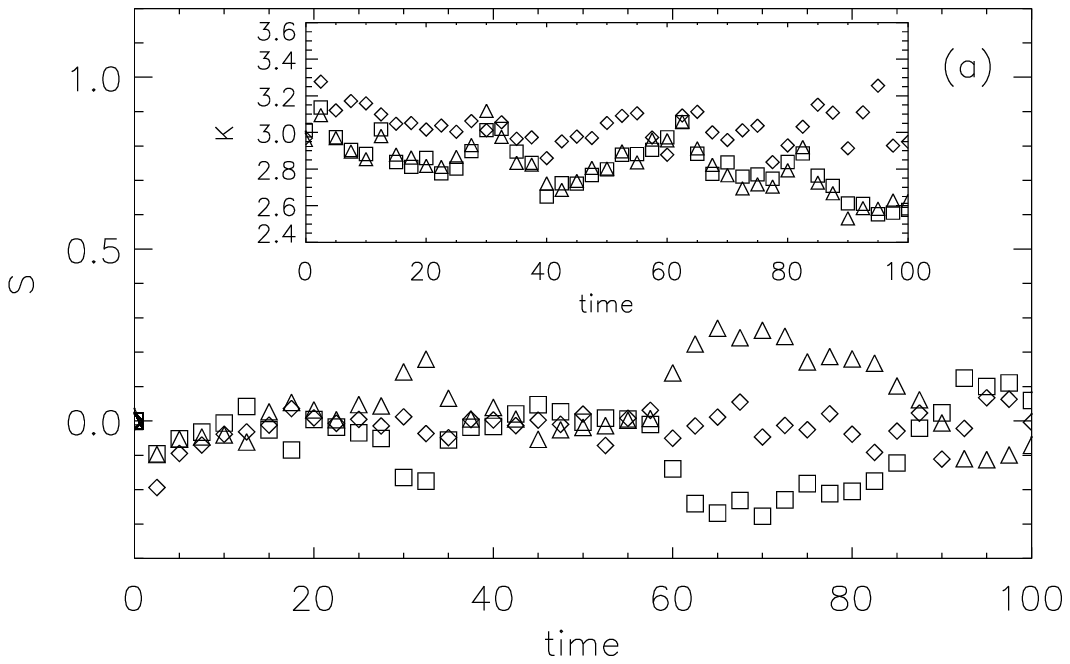}
\includegraphics[width=8cm,height=6cm]{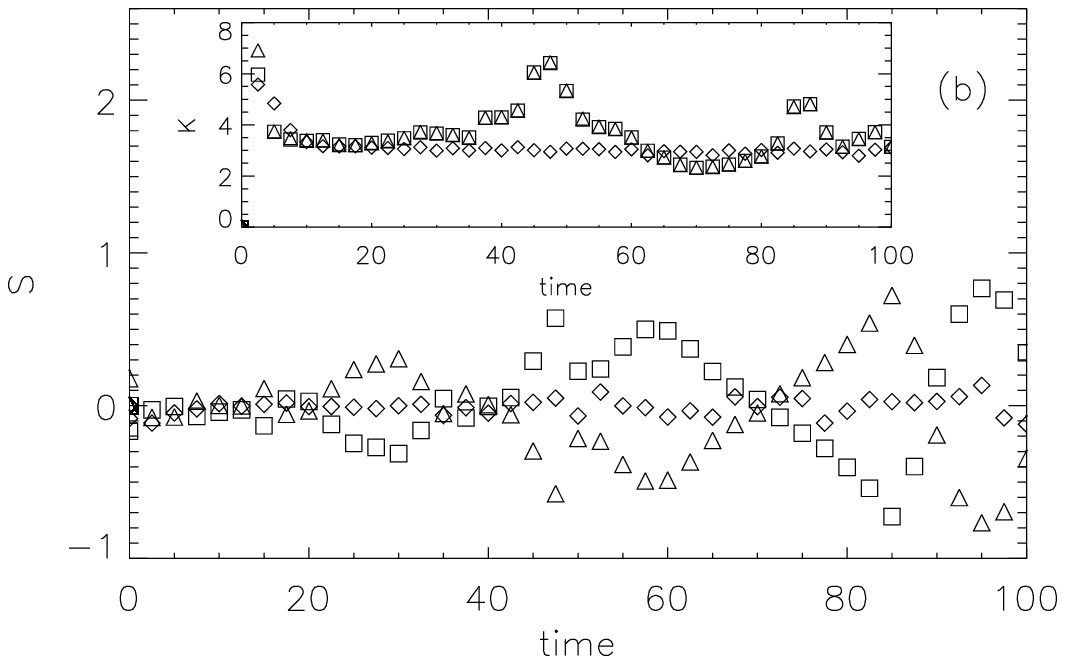}
\caption{Evolution of the velocity-derivative skewness for runs (a) L96-6 
and (b) L96H-3. Symbols are squares for $S_x$, triangles for $S_y$, and 
diamonds for $S_z$. The inset shows the evolution of velocity-derivative 
kurtosis for the same runs.}
\label{fig:skew2}
\end{figure}


\section{\label{sec:statistics}Skewness and kurtosis}

In this section we consider the time evolution of skewness and kurtosis 
of velocity derivatives in runs with and without rotation, and with and 
without helicity. The skweness $S_i$ and kurtosis $K_i$ are defined as
\begin{equation}
S_i=\left. \left< \left(\frac{\partial u_i}{\partial x_i}\right)^3 \right> 
 \middle/ 
 \left< \left(\frac{\partial u_i}{\partial x_i}\right)^2\right>^{3/2} 
 \right. ,
\end{equation}
\begin{equation}
K_i=\left. \left< \left(\frac{\partial u_i}{\partial x_i}\right)^4 \right> 
 \middle/ 
 \left< \left(\frac{\partial u_i}{\partial x_i}\right)^2\right>^{2} 
 \right. ,
\end{equation}
where $i$ denotes the Cartesian coordinates $x$, $y$, or $z$.

Figure \ref{fig:skewness} shows $S$ and $K$ for non-rotating runs 
D512-2 and L96-1 (non-helical), and D512H-2 and L96H-1 (helical). Only 
a few times are shown for the DNS runs, to compare with the LES. Overall, 
the DNS and LES show good agreement. The three components of the skewness 
fluctuate around $\approx -0.5$, a value observed in experiments 
\cite{Batchelor1949} and simulations \cite{Orszag1972}. Also, the 
kurtosis evolves towards a value near $3.5$. Helicity does not affect 
the values of $S$ nor $K$ in the absence of rotation.

When rotation is present skewness is substantially reduced, with all 
components of $S$ fluctuating around $S \approx 0$. This is shown in 
Fig.~\ref{fig:skew2} for runs L96-6 and L96H-3 (DNS show the same behavior 
and are not shown for clarity). Such a decrease of skewness with 
decreasing Rossby number has already been reported in simulations 
\cite{Cambon1997}. Anisotropy is also evident, manifested in a distinct 
behavior of $S_x$, $S_y$, and $S_z$. While fluctuations of $S_z$ are 
small, $S_x$ and $S_y$ show large and sudden departures from zero with 
$S_x \approx -S_y$ at all times. This anti-correlation between the $x$ 
and $y$ components can be qualitatively understood from the 
two-dimensionalization of the flow. For a 2D flow, the incompressibility 
condition becomes
\begin{equation}
\frac{\partial u_x}{\partial x} = - \frac{\partial u_y}{\partial y} ,
\end{equation}
which leads to $S_x \approx -S_y$.

The kurtosis in the runs with rotation has more fluctuations, but seems to 
evolves towards a value near $3$. This is clearer for $K_z$, while $K_x$ 
and $K_y$ also show signs of two-dimensionalization with $K_x \approx K_y$ 
at all times.

\begin{figure}
\includegraphics[width=8cm]{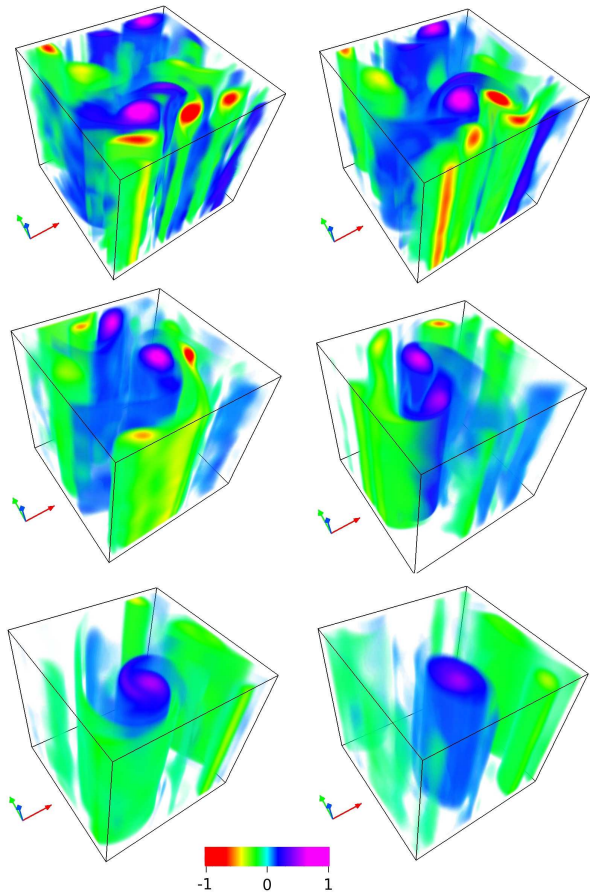}
\caption{(Color online) Visualization of $\omega_z$ at late times for 
run L96H-3. From top to bottom and from left to right, the images 
correspond to $t=42.5$, $47.5$, $55$, $70$, $t=87.5$, and $100$. Note 
that the four anti-cyclonic vortices at $t=42.5$ merge in pairs and 
become two larger columnar vortices at $t=47.5$. Eventually they merge 
again becoming one column.}
\label{fig:visualization}
\end{figure}

Visualization of the flow vorticity indicates that maxima and minima of 
$S_x$ and $S_y$ correspond to times when two column-like structures in 
the flow merge. As an example, Fig. \ref{fig:visualization} shows the 
evolution of the $z$ component of the vorticity in run L96H-3. When two 
columns with the same sign of vorticity merge, intense gradients are 
created in $u_x$ and $u_y$, giving rise to an increase or decrease in 
the values of $S_x$ and $S_y$. Columns of positive vorticity (cyclonic) 
tend to merge, while vortices of negative vorticity (anti-cyclonic) seem 
to be unstable.

\section{\label{sec:conclusions}Conclusions}

In this work we presented a study of the self-similar decay laws that 
arise in turbulent rotating flows depending on: (1) the characteristic 
scale of the initial conditions (compared with the size of the box), 
(2), the presence or absence of helicity in the flow, (3) the values of 
the Rossby and Reynolds numbers, and (4) the amount of anisotropy in 
the initial conditions. Phenomenological decay laws were obtained for 
each case considered, and the decay laws were contrasted with numerical 
results from DNS and LES using different flows as initial conditions.

A large number of power laws were identified. It is well known 
that rotation decreases the energy decay rate 
\cite{Cambon1989,Yang2004,Morize2005,Morize2006,Bokhoven2008, 
Jacquin90,Squires1994}, and our simulations are in agreement 
with this result. However, our simulations further show that in the 
presence of rotation helicity can further decrease this decay. This 
is different from the non-rotating case, where helicity does not affect 
the self-similar decay of energy. This result, together with previous 
studies in the case of forced rotating flows 
\cite{Mininni09,Mininni10a,Mininni10b} further confirm that helicity 
plays a more important role in rotating turbulence than what it does 
in the isotropic and homogeneous case.

In the presence of rotation, the decay of enstrophy is well described 
by phenomenological arguments based on isotropic scaling. This can be 
expected as enstrophy (as well as helicity) is a small-scale quantity, 
more isotropic than the energy.

The energy in rotating non-helical flows follows either a decay close 
to a $\sim t^{-1}$ law (when the integral scale of the flow is close to 
the size of the box), or a decay slightly steeper than $\sim t^{-5/7}$ 
(when the integral scale is smaller than the size of the box, and the 
large scale energy spectrum is $\sim k^4$). Better agreement with 
power-law decay is obtained when the evolution of 2D modes and 3D modes 
is considered separately. In that case, the energy in 2D modes decays 
close to $E_{2D} \sim t^{-2/3}$, and the 3D modes decay as in 
the non-rotating case, i.e., close to $E_{3D} \sim t^{-10/7}$.

These power-law decays can be obtained from phenomenological arguments 
that consider the energy in 2D and 3D modes separately, that assume 
approximately constant axisymmetric integrals instead of the isotropic 
Loitsyanski's integral, and that take into account the slow down in 
the energy transfer associated with rotation. Note we are not 
claiming there is decoupling between 2D and 3D modes in rotating flows, 
a topic which is beyond the scope of this work. What we show instead 
is that the energy in 2D and 3D modes in the simulations decay with 
different rates, both following power laws, and that considering this 
in the phenomenological description gives better agreement with the 
numerical results.

The decay of energy in the presence of rotation and helicity shows 
further variety. When the integral scale of the flow is close to 
the size of the box, the energy decay is close to $E \sim t^{-1/3}$. 
This decay can be obtained from phenomenological arguments taking 
into account the role played by the helicity cascade in further 
slowing down the energy transfer. In simulations with integral scale 
smaller than the size of the box, the decay is between 
$E \sim t^{-5/7}$ and $\sim t^{-1/3}$.  As in the non-helical case, 
clearer power laws arise if the decay of $E_{2D}$ and $E_{3D}$ 
is considered. In that case, $E_{2D}$ shows decays between 
$\sim t^{-1/2}$ and $\sim t^{-1/6}$, and $E_{3D}$ shows decays close 
to either $E \sim t^{-2}$ or $E \sim t^{-10/7}$. 

The results with helicity seem to be more dependent on scale separation 
(i.e., on the separation between the initial integral scale of the flow, 
and the size of the box), and on initial anisotropy. It is worth 
mentioning that the importance of the initial conditions in the decay 
of rotating turbulence has been recently pointed out also in experiments 
\cite{Moisy}.

Finally, we presented a study of the time evolution of the skewness and 
kurtosis of velocity derivatives. Two-dimensionalization of rotating flows 
leads to an anti-correlation of the $x$ and $y$ components of the 
skewness, which fluctuate around zero. Large departures of these quantities 
from this value are associated with merging events of columns in the flow.

\begin{acknowledgments}
The authors acknowledge fruitful discussions with A. Pouquet, and help 
implementing the LES from D. Rosenberg and J. Baerenzung.
Computer time was provided by NCAR and CeCAR. TT and PDM acknowledge 
support from UBACYT Grant No. 20020090200692, PICT Grant No. 2007-02211, 
and PIP Grant No. 11220090100825.
\end{acknowledgments}


\end{document}